\documentclass[fleqn,usenatbib]{rasti}

\usepackage{newtxtext,newtxmath}

\usepackage[T1]{fontenc}

\DeclareRobustCommand{\VAN}[3]{#2}
\let\VANthebibliography\thebibliography
\def\thebibliography{\DeclareRobustCommand{\VAN}[3]{##3}\VANthebibliography}

\usepackage{graphicx}	
\usepackage{mathtools}

\usepackage{subcaption}
\usepackage{textcomp, gensymb}
\usepackage{upgreek}
\usepackage{anyfontsize}

\usepackage{float}
\floatstyle{plaintop}
\restylefloat{table}

\usepackage{changepage}                 
\newlength{\offsetpage}
{\end{adjustwidth}}

\usepackage[usenames,dvipsnames]{xcolor} 

\renewcommand\fbox{\fcolorbox{red}{white}}

\title[Satellite Classification Using Stellar Occultations]{Classification of LEO Satellites Using Occultations of Background Stars}

\author[B. F. Cooke et al.]{
Benjamin F. Cooke,$^{1,2}$\thanks{E-mail: benjamin.cooke@warwick.ac.uk}
Don Pollacco,$^{1,2}$
James A. Blake,$^{1,2}$
Paul Chote,$^{1,2}$
Stuart Eves,$^{3}$
Will Feline$^{4}$
\newauthor
and
Grant Privett$^{4}$
\\\\
$^{1}$Centre for Space Domain Awareness, University of Warwick, UK\\
$^{2}$Department of Physics, University of Warwick, UK\\
$^{3}$SJE Space Ltd, UK\\
$^{4}$Defence Science and Technology Laboratory, UK
}

\date{Accepted XXX. Received YYY; in original form ZZZ}

\pubyear{\the\year{}}

\begin{document}
\label{firstpage}
\pagerange{\pageref{firstpage}--\pageref{lastpage}}
\maketitle

\begin{abstract}
We present the result of a proof-of-concept simulation designed to classify LEO satellites based on their occultations of background stars. We generate satellite shapes drawn from two broad shape classes, `boxwing' and `square'. We then simulate the resulting occultation photometry that would be caused by these satellites orbiting in LEO and intersecting with background stars. The resulting data is then inverted to attempt to recover the input shape and classify the satellite correctly. We find that the technique is theoretically sound, but ambitious with current telescope capabilities. We construct an equation for the required success rate of the method, as a function of exposure time and density of background stars. We find that successful classification requires short exposure times and high background stellar densities. For success rates in excess of 75\%, we find a required exposure time of $\sim2.0\times10^{-3}$\,s, and $\sim500$ stars along the satellites path. Results are presented in terms of these two key parameters, and are discussed in the context of current observational capabilities and alternative satellite characterisation methods.
\end{abstract}




\section{Introduction}
\label{sec:Introduction}


Observing Resident Space Objects (RSOs) in Low Earth Orbit (LEO) is a rapidly expanding field. The number of objects in LEO, both active and inactive satellites \citep{2020AdSpR..65..351O}, alongside rocket bodies and various debris \citep{2023AcAau.210..465P} is increasing exponentially, and will continue to do so \citep{2019JSSE....6...80M,2020Aeros...7..133C,2022A&G....63.2.14B}. Observations at LEO are usually carried out using radio observations \citep{10.1093/mnras/stv1010, MUCIACCIA2024143}, but in recent years optical observations have increased due to improvements in technology and reduction in cost. A key change is the adoption of sCMOS devices over the more traditional CCD-based approach \citep{COOKE_2023,2023AdSpR..72.2064Z}. Both techniques can be used for object detection, but also for characterisation of known objects \citep{2023GPSS...27..178S,2023Senso..23.9668S}. At radio wavelengths, the Radar Cross Section (RCS) can allow for inferences to be made regarding target size and material \citep{EMERY2017291}. Likewise, optical observations allow for some elements of characterisation as well. Single colour observations can suggest target size (under some assumptions) but multi-wavelength (multi-colour) photometry can allow for further characterisation including target material and age \citep{ZHAO20162269,10.1093/mnras/stae693,AIREY20255757}.

Both of these techniques have disadvantages when it comes to observing particular LEO targets. Radio observations are dependent on RCS, as opposed to true size, thus the use of particular materials and designs can limit the reliability of radio-based characterisation \citep{2019LPICo2109.6164X}. Traditional optical observations rely on the reflection of sunlight off the target and into the detector. Again, this means that observations can be limited by attempts to reduce the reflectivity of a satellite \citep{2020A&A...637L...1T,2021A&A...647A..54T}. Additionally, optical observations also require that a target is sunlit, yet must also be carried out during night time hours, limiting observations to the twilight hours, depending on the target orbit height \citep{COOKE_2024}.

A further method for satellite identification and characterisation is to utilise directly resolved observations. Due to the small angular size of LEO satellites (a 1\,m satellite at zenith, with an altitude of 1000\,km subtends an angle of $\sim0.2\arcsec$) and the relatively high velocities, resolved observations are challenging and require specialised techniques. Optically, this can be done using Multi-Frame Blind Deconvolution (MFBD) to mitigate atmosphere-induced observation degradation \citep{1993JOSAA..10.1064S}. This technique assumes the PSF is convolved with the true target image and attempts to solve both concurrently \citep[e.g.,][]{2019amos.confE...5W}. An alternative approach is to use Adaptive Optics (AO) to mitigate scintillation effects \citep{TOKUNAGA20141089}. Using a known star, wavefront distortions can be measured and adjusted for using a deformable mirror. The adjustments can then lead to improved resolution of satellite images \citep[e.g.,][]{petit:hal-03051923,2023amos.conf..148M}. For radio observations, Inverse Synthetic Aperture Radar (ISAR) can be used to resolve a satellite as it passes overhead \citep{MARTORELLA2014987}. The motion of the target with respect to the observer gives a temporal element to the radio reflectivity and thus can allow for the resolution of satellite images \citep[e.g.,][]{2018SPIE10633E..0LA,2023amos.conf...34J}. Both of the above techniques require sophisticated instrumentation and processing, and neither are immune to the limitations already mentioned.

An observational technique designed to subvert these disadvantages is to use stellar occultations \citep{2017amos.confE..80S}. As an RSO moves across the sky, it temporarily occults background sources and these occultations can reveal details about the target. The benefit of relying on these occultations is that they are entirely affected by the outline of the object, i.e. the size and shape \citep{2021ascl.soft08025G}. Attempts to reduce the RCS or reflectivity of the object will have no effect on its occultations of background sources \citep{2022A&A...667A..45G}. Additionally, since the observations do not require the object to be in sunlight, observations can be carried out when traditional optical observations are inaccessible. This means the technique is potentially very powerful, able to be resistant to even deliberate attempts to make an object less detectable while also being less restricted in time. It is notable however, that the technique is not available while the target is sunlit, as the reflected light from the target would obscure the occultations.

The duration and depth of individual occultations is affected by the physical size of the occulting object and the speed of its motion. The faster the motion and the smaller the target size, the shorter any occultation may be (additionally, for a non-circular object, the part of the object which passes in front of the star also affects the occultation time). Any occultation results in a dip in stellar flux, with the ratio of occultation duration to exposure time determining the depth of any flux drop \citep{2021ascl.soft08025G}. The number of separate occultation events is influenced by the size of the occulting object but is also dependent on the density of stars in the field of view (FoV). Examining the occultation data, with knowledge of the exposure time used and the background stellar field, can allow us to attempt to recover the size and shape of the occulting object \citep{2022A&A...667A..45G}.

In the following manuscript we utilise this occultation technique and attempt to define the necessary requirements to use the technique to classify LEO satellites of known orbit but unknown shape and size. We begin with a description of the technique (Sect. \ref{sec:Technique}) and set out our simulation parameters (Sect. \ref{sec:Simulation parameters}). We then describe the generation of our data (Sect. \ref{sec:Photometry generation}), before detailing our method of recovering the satellite classes (Sect. \ref{sec:Shape recovery}). We then present our results (Sect. \ref{sec:Results}) and discuss our conclusions (Sect. \ref{sec:Discussion and conclusions}).

\section{Methodology}
\label{sec:Methodology}

\subsection{Technique}
\label{sec:Technique}


To determine the feasibility of the method for classification of LEO satellites we conduct a simulation, testing our method on a large sample of theoretical satellites. Since this is a new and ambitious technique, a simulation-based approach is pertinent to avoid the waste of time and telescope resources before a reasonable limiting case is known. We generate a range of theoretical satellite shapes drawn from two distinct classes, `boxwing' and `square'. Boxwing satellites are symmetrical, with a central, rectangular body, and two extended wings, whereas square satellites are comparable, but without the extended wing structures. Simulated satellites are assumed to be fully opaque. We then simulate them as crossing over an area of sky with a known stellar distribution at LEO altitudes and speeds. We generate the expected occultation photometry based on a known exposure time and then attempt to recover the input shape of the targets, classifying results into two groups. Comparing the success rate of this method as a function of various simulation parameters will allow us to predict the limiting conditions required to utilise this method in reality.

\subsection{Simulation parameters}
\label{sec:Simulation parameters}

\begin{figure}
    \centering
    \includegraphics[width=\columnwidth]{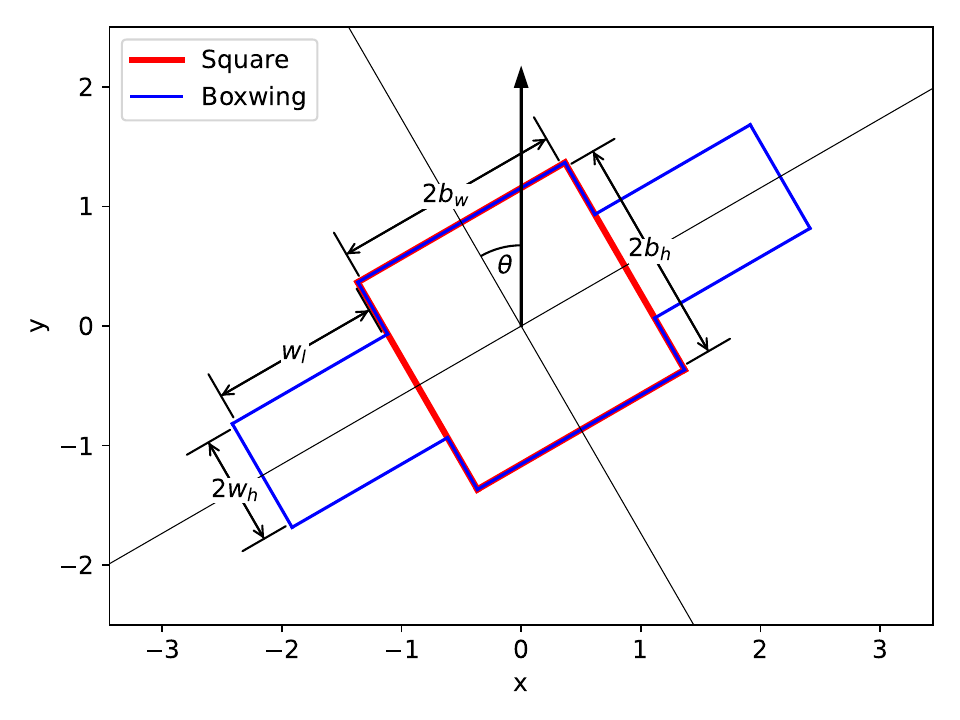}
    \caption{Geometry of the two classes of satellite shape showing the relevant shape parameters. A square satellite (red) is defined by two parameters ($b_w$ and $b_h$) and a boxwing satellite (blue) is defined by four parameters ($b_w$, $b_h$, $w_l$ and $w_h$). The arrow denotes the direction of motion and $\theta$ the rotation angle (here taken as $30\degree$). Thin grey lines show the shape axes.}
    \label{fig:shape_geometry}
\end{figure}

Our simulated satellites are split evenly between the two shape classifications, boxwing and square. We select a satellite body width ($b_w$) drawn from a uniform distribution $\mathcal{U}[0.05,2.0]$. The satellite body height ($b_h$) is then scaled from this width value, with the scaling factor drawn from a distribution $\mathcal{U}[0.75,1.25]$. $b_w$ and $b_h$ are measured from the vertical and horizontal body axes respectively. For a square satellite this is all that is required, but for a boxwing we also require two wing parameters. First the wing length ($w_l$, measured from the edge of the body), again, scaled off the body width, with the scaling factor drawn from $\mathcal{U}[0.5,2.0]$ and secondly the wing height ($w_h$, measured from the horizontal body axis), scaled from the body height, with a scale factor drawn from $\mathcal{U}[0.25,0.75]$. The wings are positioned as extending along the body width dimension, located centrally along the body height. To convert relative sizes to actual dimensions we multiply all dimensions by 1\,m. Finally, we generate an angle by which the satellite shape is rotated before the simulation begins ($\theta$). The angle is in radians, drawn from $\mathcal{U}[0,\pi]$. In 3D space, this corresponds to a rotation around the axis connecting the satellite to the observer. For this proof-of-concept work we consider only a single axis of rotation, but further work could consider a rotation around any or all of the three axes. Figure \ref{fig:shape_geometry} shows a representation of the geometry of the two classes of shape and the relevant shape parameters.

For a LEO satellite we assume an orbital height above the Earth of 1000\,km, corresponding to the centre of the LEO region, generally defined to end at 2000\,km \citep{LEO} (the number of current LEO targets actually peaks slightly below this, at $\sim800$\,km, however this difference will have little effect on our results thus, for generality, we use the centre of the range). Under the assumption of Keplerian motion, this gives us an orbital angular speed of 0.06\,deg/s and a velocity of approximately 7.5\,km/s (the equivalent velocity at GEO, with a height of 36\,800\,km, is approximately 3\,km/s). From the observers point of view, when observed at zenith, the observed rate of motion is larger, approximately 0.43\,deg/s. Observing closer to the horizon decreases the effective rate as the satellite-observer distance is increased.

Once the satellite shape has been chosen, we define the background stellar field. Based on the maximum potential extent of a satellite (from the shape distributions discussed above) and the velocity and direction of the satellite, we use an area of space defined by a circle with diameter equal to the maximum satellite dimension moving along a line which has length equal to the distance moved by the satellite over 1\,s. This 1\,s value is chosen to give a short arc over which a satellite will not change orientation and which will fit into the FoV of most telescopes. The arc length, and thus area, can be scaled later (see Sect. \ref{sec:Discussion and conclusions}). This area is then randomly populated by $n$ point-like stars, where $n$ is drawn from a uniform distribution $\mathcal{U}[0,1000]$. For a target with orbital velocity 7.5\,km/s, observed at zenith, this corresponds to a density of approximately $n/0.53$ stars per arcmin squared. Observing closer to the horizon increases the effective density, whereas observing for a longer window reduces it. 
Finally, we choose the exposure time. Based on the satellite sizes and velocities described above, we expect a maximum occultation duration of $\sim2\times10^{-3}$\,s. We select exposure times around this range, testing both shorter and longer exposures. Thus we draw the exposure time, in seconds, from $\mathcal{U}[0.0004,0.004]$. From these parameters we then generate the expected photometry.

\subsection{Photometry generation}
\label{sec:Photometry generation}



\begin{figure}
    \centering
    \includegraphics[width=\columnwidth]{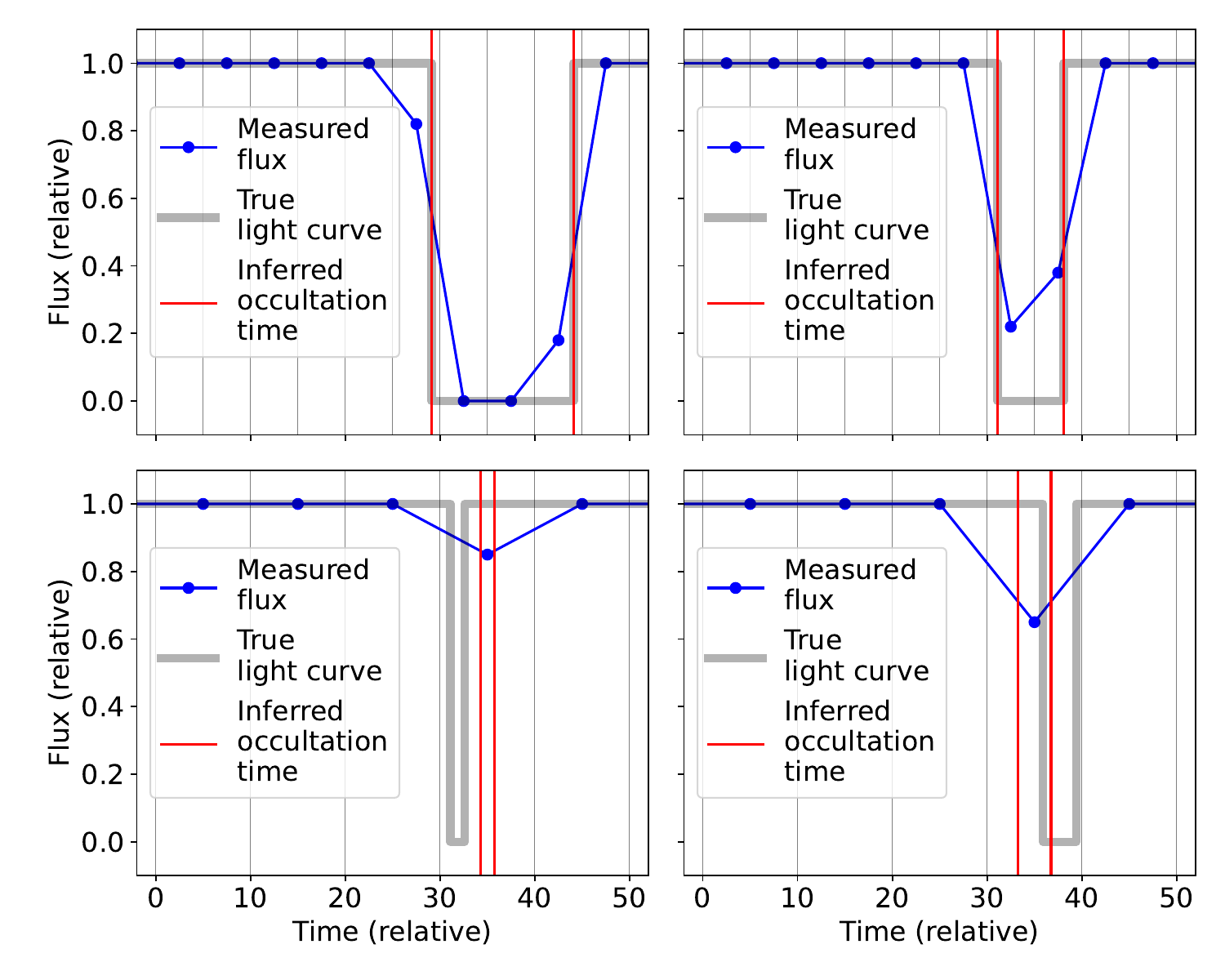}
    \caption{Idealised light curves from occultation events. Blue data shows the measured flux, the thick grey line shows the true light curve and thin grey lines mark the boundaries between individual exposures. The vertical red lines show the start and end of the occultation as inferred from the data. Top row panels show exactly recovered occultation times (i.e. fully resolved occultations), bottom row panels show approximately recovered occultation times.}
    \label{fig:lightcurves}
\end{figure}

To simulate photometry we move the simulated satellite shape through the stellar field, recording if and when it intersects with any of the simulated stars with an intersection being defined as a point at which a star aligns with any of the edges of the satellite shape. Intersections thus occur in pairs, denoting when a star begins to be occulted by the satellite and when it reappears. We allow for multiple distinct occultations of a star by the satellite, accounting for separate occultations by the body and wing of the satellite depending on the angle of rotation and shape dimensions. Each pair of intersections causes an occultation feature in the simulated photometry. Due to the point-like nature of the stars, the occultation is total, i.e., if observed at exactly that point in time, the star would be completely occulted and thus undetectable. However, due to the non-zero exposure time of observations, the occultation appears as a dimming of the star. Should the occultation be entirely contained within a single exposure we can measure only its duration, with the occultation duration divided by the exposure time being equal to the reduced flux divided by the total flux. The exact time of the occultation is unknown, save that it occurs within the relevant exposure time window, thus we assume it to happen centrally. If however, the occultation occurs across the boundary of two exposure times we can measure the start and end time of the occultation directly, by measuring how long the occultation lasts in each exposure time window and subtracting this from either the start or end of the exposure time window. Figure \ref{fig:lightcurves} shows some model light curves obtained from stars being occulted. In each panel the blue data shows the measured photometry, the thick grey line shows the true light curve and the thin grey lines mark the boundaries between individual exposures. The vertical red lines show the start and end of the occultation as inferred from the data. In the top two panels, the occultation crosses one or more exposure time boundaries and thus the occultation time is recovered exactly (i.e. is fully resolved). In the bottom two panels the occultation is contained within a single exposure and thus only the occultation duration is known, its exact location within the exposure is unknown (and assumed to be central).

Since the position of the satellite as a function of time is known, we can take the time and position of each intersection point (exact or approximated) and plot them relative to the centre of the satellite. The intersection points will then approximate the outline of the satellite, with exactly inferred intersection times matching up with the edges of the satellite shape, and approximated times being slightly offset along the direction of motion. This arrangement of intersection points can then be used to attempt to recover the original satellite shape.

\subsection{Shape recovery}
\label{sec:Shape recovery}

\begin{figure}
    \centering
    \includegraphics[width=\columnwidth]{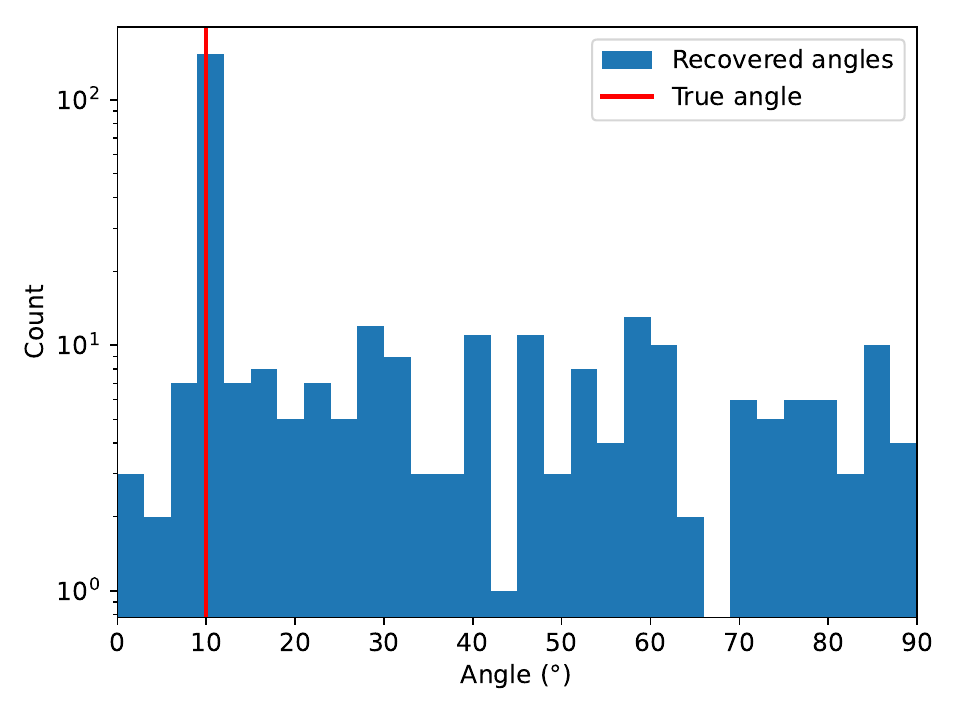}
    \caption{Distribution of recovered angles for a test case with a true rotation of $10\degree$. The true rotation value is marked in red and is well recovered as the modal peak in the distribution.}
    \label{fig:angle_hist}
\end{figure}

The first step to recovering the satellite shape is to attempt to recover the rotation angle. To do this we consider the angle between pairs of intersection points. Pairs of intersection points define a line, and the angle is measured between this line and the positive x-coordinate axis of the frame. Where both intersection points align with the edges of the satellite, the angle between them will be equal to the rotation angle, under the assumption of rotation about a single axis ($\pm90\degree$ where the points are on adjoining edges). Points not exactly aligning with shape edges will have various angles from each other, thus the most common angle between intersection points should be equal to the rotation angle of the satellite shape. Figure \ref{fig:angle_hist} shows the distribution of recovered angles for a test case with a true rotation angle of $10\degree$. The true angle is marked in red, and is well recovered as the modal peak in the distribution. This approach is valid for the satellite shape we use here, since they are composed entirely of straight line segments. For more complex satellite shapes this approach must be further developed.

\begin{figure*}
    \centering
    \begin{subfigure}{\columnwidth}
        \centering
        \includegraphics[width=\columnwidth]{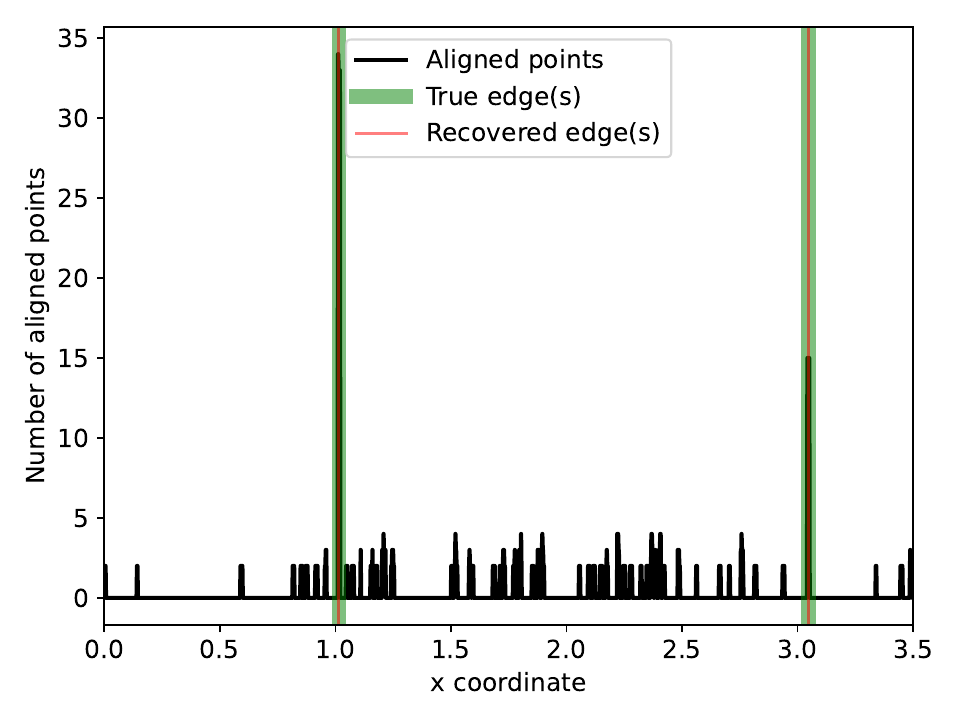}
        \caption{Boxwing fit, perpendicular direction}
        \label{fig:hist_boxwing_x}
    \end{subfigure}
    \begin{subfigure}{\columnwidth}
        \centering
        \includegraphics[width=\columnwidth]{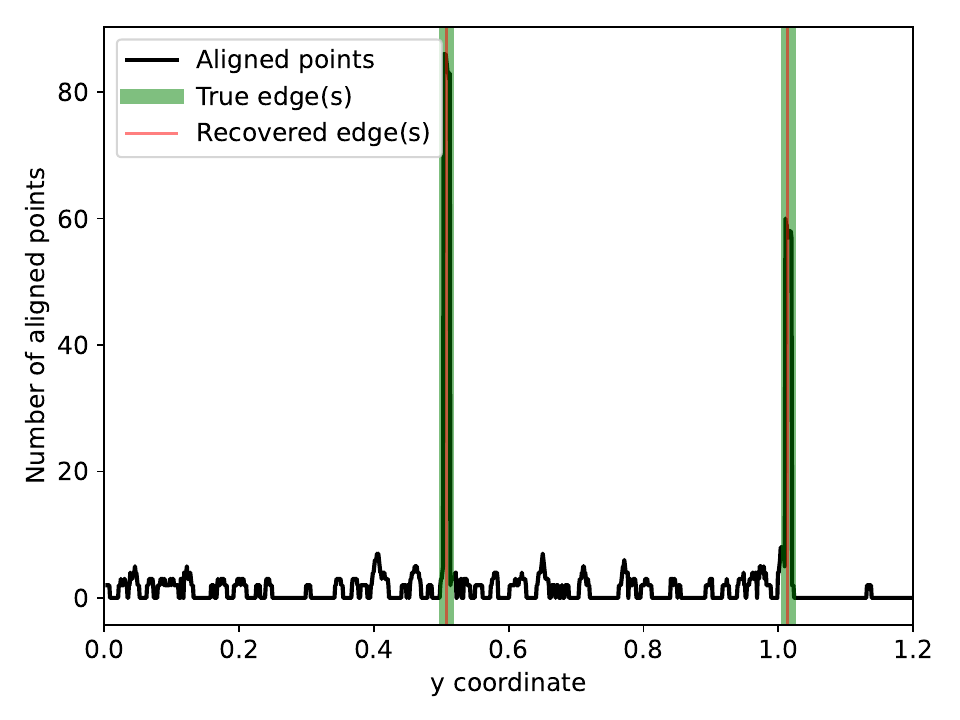}
        \caption{Boxwing fit, parallel direction}
        \label{fig:hist_boxwing_y}
    \end{subfigure}
    \begin{subfigure}{\columnwidth}
        \centering
        \includegraphics[width=\columnwidth]{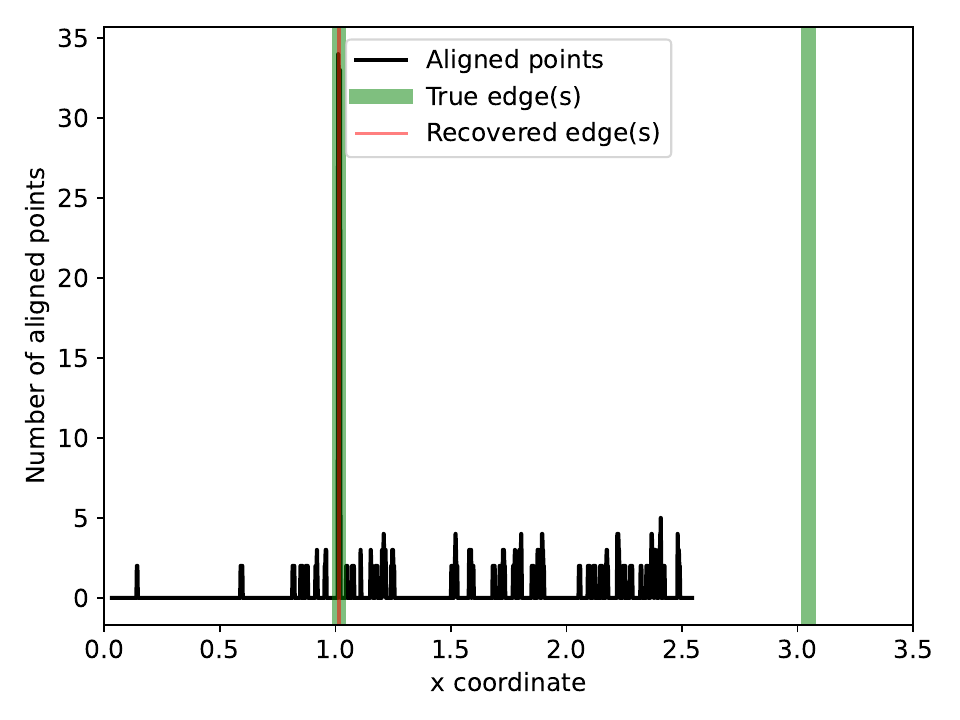}
        \caption{Square fit, perpendicular direction}
        \label{fig:hist_square_x}
    \end{subfigure}
    \begin{subfigure}{\columnwidth}
        \centering
        \includegraphics[width=\columnwidth]{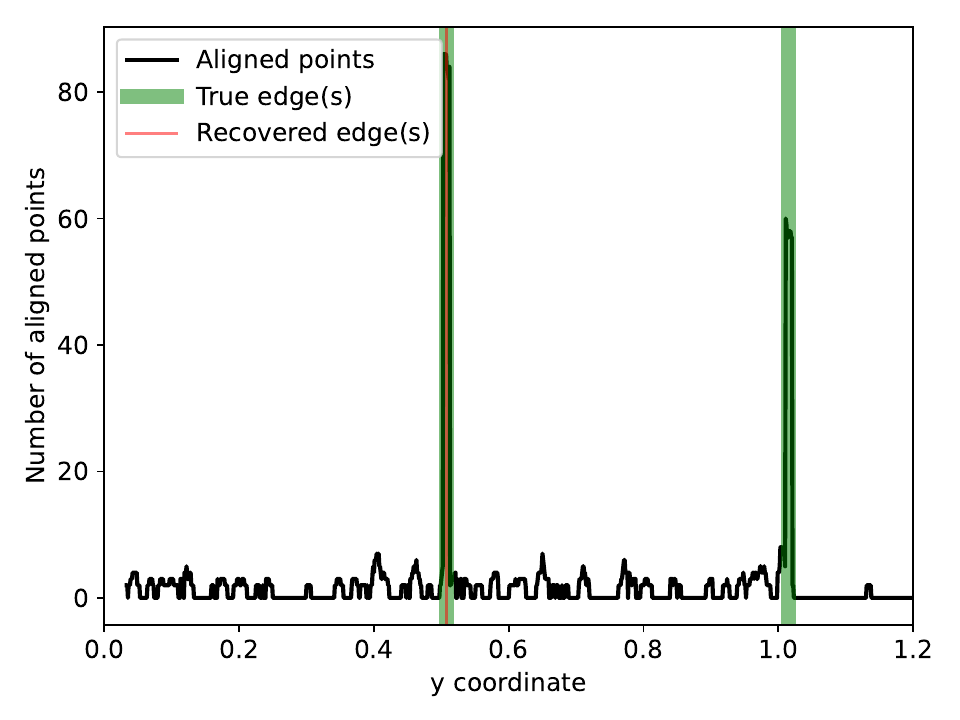}
        \caption{Square fit, parallel direction}
        \label{fig:hist_square_y}
    \end{subfigure}
    \caption{Satellite edge recovery for a boxwing satellite. We show plots of the number of aligned intersection points for a range of test edges (defined by the coordinate at which they cross the axis). For each attempted shape fit, we search for edges parallel and perpendicular to the rotation angle. The green lines show the location of the true edges (two per direction, since the true shape is a boxwing), and the red lines, the best recovered edges. For a boxwing fit, we select the best two edges in each direction, for square we only select the single best edge in each direction.}
    \label{fig:hists}
\end{figure*}

With the angle recovered we then attempt to determine the position of the satellite shape edges. We position a sequence of lines with angle equal to the recovered rotation angle, spread out perpendicularly. We then count how many intersection points are sufficiently close to each line, giving a plot of tested edge versus number of aligned points. Making the assumption that the satellite is symmetrical means that we only have to test edges along one side of the satellite shape, which is then reflected to get the opposite side edge. When the tested edge aligns with a true satellite edge, the number of close intersection points should show a peak. Peaks of sufficient height are then taken as potential satellite edges. The process is repeated with an angle $90\degree$ offset from the recovered angle to determine the position of the perpendicular satellite edges. The tested edge is denoted by the coordinate at which it crosses the x/y axis for perpendicular/parallel edges respectively (the x/y coordinate system is aligned with the FoV and has its origin at the satellite centre). We are therefore left with a set of potential edges, parallel and perpendicular to the satellite rotation angle. Figure \ref{fig:hists} shows the edge test plots for a test-case satellite (in this case, the true satellite shape is a boxwing, thus there are two true edges in each direction). For this example, the parallel direction probes $b_h$ and $w_h$ and the perpendicular direction probes $b_w$ and $w_l$, as shown in Figure \ref{fig:shape_geometry}. Since the satellite is a boxwing, and thus wider than it is tall, the maximum perpendicular recovered edges are further from the origin. We show the distribution of aligned intersection points for all tested edges, giving two plots for each fit, in perpendicular directions. The green lines are the location of the true edge(s) and the red lines are the recovered edge(s). For the boxwing fits, two edges are recovered in each direction, and for the square fits, only one edge is recovered. Because square satellites have a smaller maximum width, when attempting to fit a square we test a smaller range of coordinates, thus, in Figure \ref{fig:hist_square_x}, the tested coordinates end around 2.5, before the second edge of the true boxwing shape. When fitting a boxwing shape, both true edges are recovered in each direction, and when fitting a square shape, one true edge is recovered in each direction.

\begin{figure}
    \centering
    \begin{subfigure}{\columnwidth}
        \centering
        \includegraphics[width=\columnwidth]{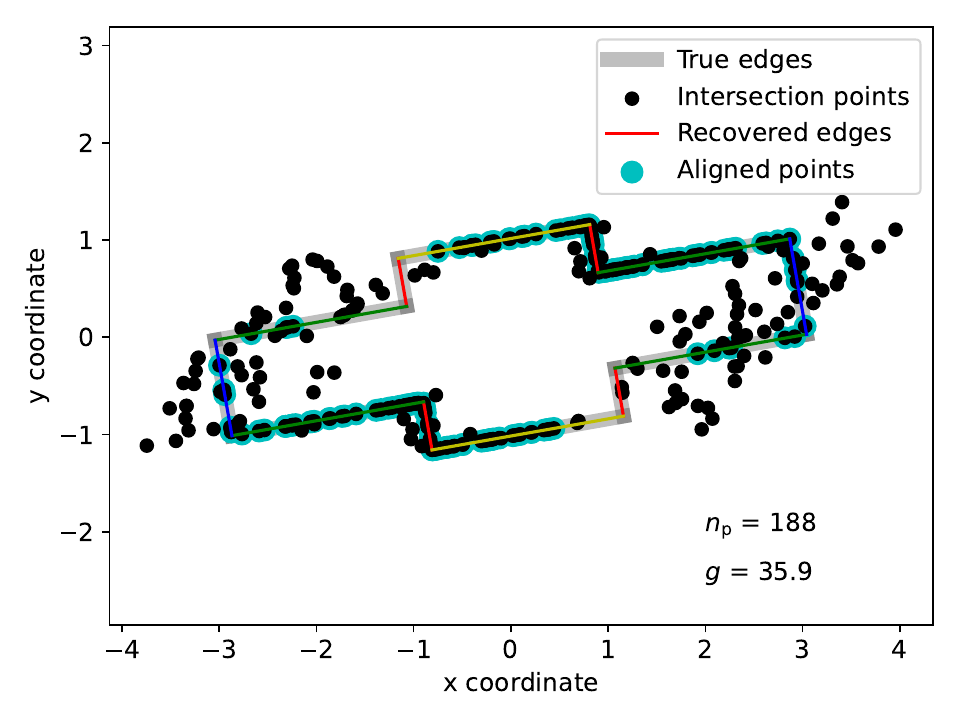}
        \caption{Best fit boxwing shape}
        \label{fig:best_fit_boxwing}
    \end{subfigure}
    \begin{subfigure}{\columnwidth}
        \centering
        \includegraphics[width=\columnwidth]{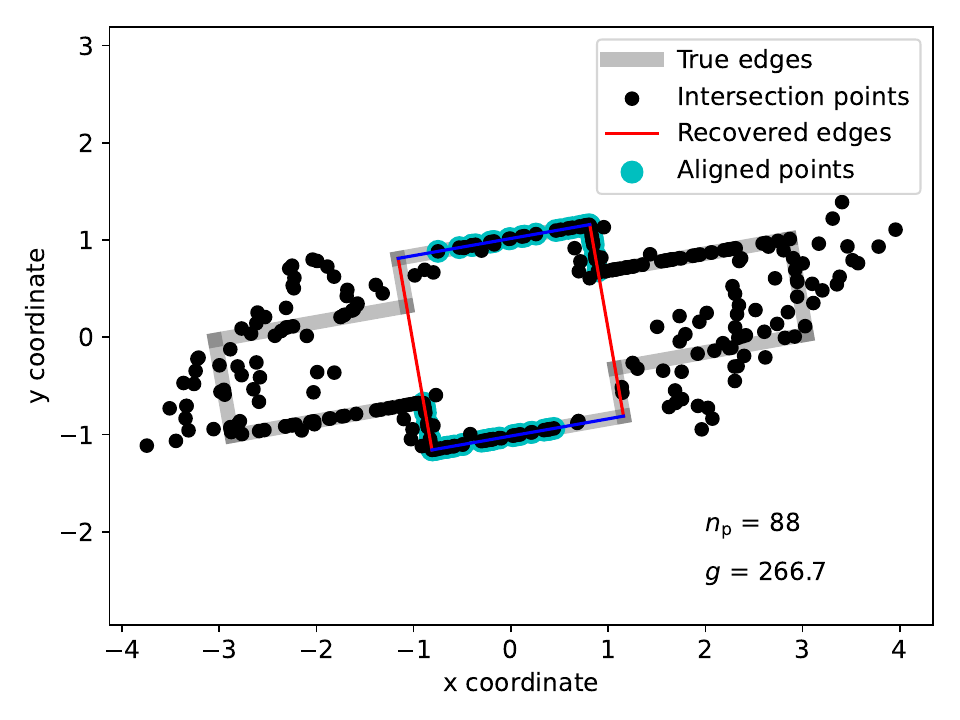}
        \caption{Best fit square shape}
        \label{fig:best_fit_square}
    \end{subfigure}
    \caption{Best fit shapes. Plots showing the best fit recovered shape for both a boxwing and square shape (in this case, the true shape is a boxwing). The true edges are in grey and the intersection points are in black. The coloured lines are the recovered edges (edges reflected across the origin are the same colour). The number of aligned points, $n_{\rm p}$, and the goodness-of-fit measure, $g$, for each fit are also given. Aligned intersection points are highlighted in cyan. The boxwing fit is preferred in this case since it has a larger number of aligned points. Additionally, it has a lower $g$ value so would be preferred even if the number of aligned points were equal between the two fits.}
    \label{fig:best_fits}
\end{figure}

We then attempt to fit two separate satellite shapes, a square and a boxwing. For the square, we take each pair of potential edges (one parallel and one perpendicular) and draw a square shape using the four edges (each recovered edge is reflected about the satellite centre). For each drawn shape, we check that the ratio of width to length is within the allowed range (as described in Sect. \ref{sec:Simulation parameters}). For an expanded simulation, where the shape ratio limit may be less well constrained, ratio limits could be inferred from a sigma clipped distribution of all recovered shape ratios. For the scenario described here however, we can use the absolute truth. We then consider all the recovered intersection points, and determine their minimum distance from a proposed satellite shape edge. Additionally we define a minimum distance from an edge required to classify an intersection point as exactly along one of the fitted edges. To compare multiple square fits we first rank the fits by the number of points aligning exactly along one of the proposed shape edges, $n_{\rm p}$. If two fits have the same number of exactly aligned points, we take the sum of the distance of all intersection points from their closest edge, defined as the goodness-of-fit, $g$. The fit having the smallest $g$ value is selected. The process is repeated with a boxwing fit, the difference being that a boxwing fit requires two parallel edges and two perpendicular edges. For each set of four edges we infer the boxwing shape outline, and the shape is then tested as described above, once again, first checking that the ratios between body width/length and wing width/length are within the allowed ranges. This process leaves us with the best fitting square and the best fitting boxwing shape. Figure \ref{fig:best_fits} show the best fitting boxwing and square fits for our test-case (a boxwing satellite), based on the results from Figure \ref{fig:hists}. The grey lines are the true edges, the black points are the intersection points (highlighted cyan when aligned with a true edge) and the coloured lines are the recovered edges (edges reflected across the origin are the same colour). Again, the x/y coordinate system is aligned with the FoV and has its origin at the satellite centre.

From Figure \ref{fig:best_fits} we see that the intersection points are either tightly constrained along the true edges of the satellite, or have a noticeable scatter. The aligned points are a result of occultations which are exactly recovered (Figure \ref{fig:lightcurves}, top row) whereas the points with more scatter are a result of occultations which are not exactly recovered (Figure \ref{fig:lightcurves}, bottom row). The remainder of this paper considers all data together (though aligned points are more highly weighted as described above). As an alternative, Appendix \ref{sec:Resolved occultations only} considers the analysis and recovery if only fully resolved occultations are considered.

The best fitting square and boxwing fits are then compared with each other, and the fit with either the largest number of aligned points, $n_{\rm p}$, or the smallest total distance of intersection points from edges, $g$, (when the number of aligned points is equal) is chosen. The recovered shape is then classified as either boxwing or square. In the case that one of the shapes does not return a valid fit, i.e. if the boxwing fit fails to return two peaks in each direction, or the parallel and perpendicular edges are not within the allowed ratio range, then that fit is rejected and the other shape is selected automatically. In the case that neither shape type returns a valid fit, the data is regarded as too limited to make a robust fit and is classified as a rejection. The result is that each satellite is classified into one of three categories; square, boxwing, or rejection. Note that rejection is by definition incorrect, but is preferable to simply returning a shape at random.

\subsection{Full simulation}
\label{sec:Full simulation}


For the full simulation we generate 100\,000 satellites, split evenly between boxwing and square shapes. For each, we generate a random shape and angle as described in Sect. \ref{sec:Simulation parameters} and choose a value of $n$ and $\Delta t$ drawn from $\mathcal{U}[0,1000]$ and $\mathcal{U}[0.0004,0.004]$ respectively. We then simulate the resulting photometry, and attempt to recover the original shape, classifying the satellite into one of three groups, as described in Sects. \ref{sec:Photometry generation} and \ref{sec:Shape recovery}. The results can then be analysed in terms of the full population of targets and be explored as a function of individual parameters.

\section{Results}
\label{sec:Results}

\begin{figure}
    \centering
    \begin{subfigure}{\columnwidth}
        \centering
        \includegraphics[width=\columnwidth]{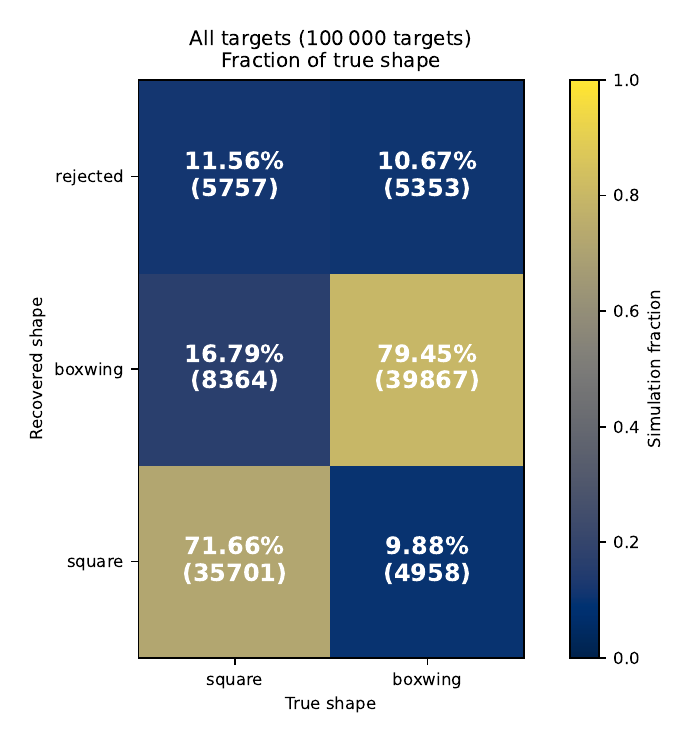}
        \caption{Including satellites with no successful fits}
        \label{fig:accuracy_plot1a}
    \end{subfigure}
    \begin{subfigure}{\columnwidth}
        \centering
        \includegraphics[width=\columnwidth]{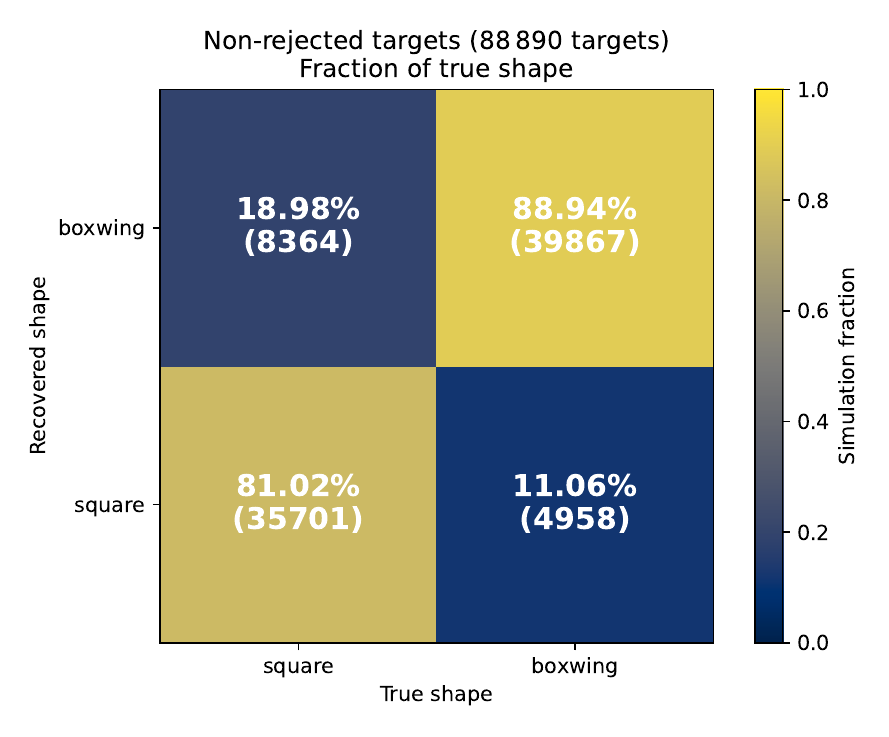}
        \caption{Only including satellites with at least one successful fit}
        \label{fig:accuracy_plot1b}
    \end{subfigure}
    \caption{Satellites separated by their true shape and their shape as returned by our simulation. Percentages are given as a fraction of the number of satellites of each true classification (i.e. totals are summed vertically).}
    \label{fig:accuracy_plot1}
\end{figure}
\begin{figure}
    \centering
    \begin{subfigure}{\columnwidth}
        \centering
        \includegraphics[width=\columnwidth]{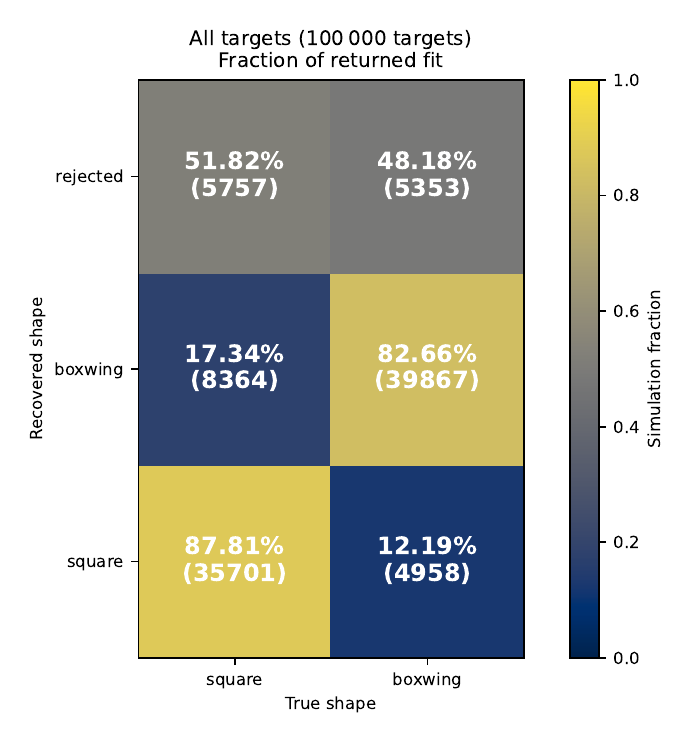}
        \caption{Including satellites with no successful fits}
        \label{fig:accuracy_plot2a}
    \end{subfigure}
    \begin{subfigure}{\columnwidth}
        \centering
        \includegraphics[width=\columnwidth]{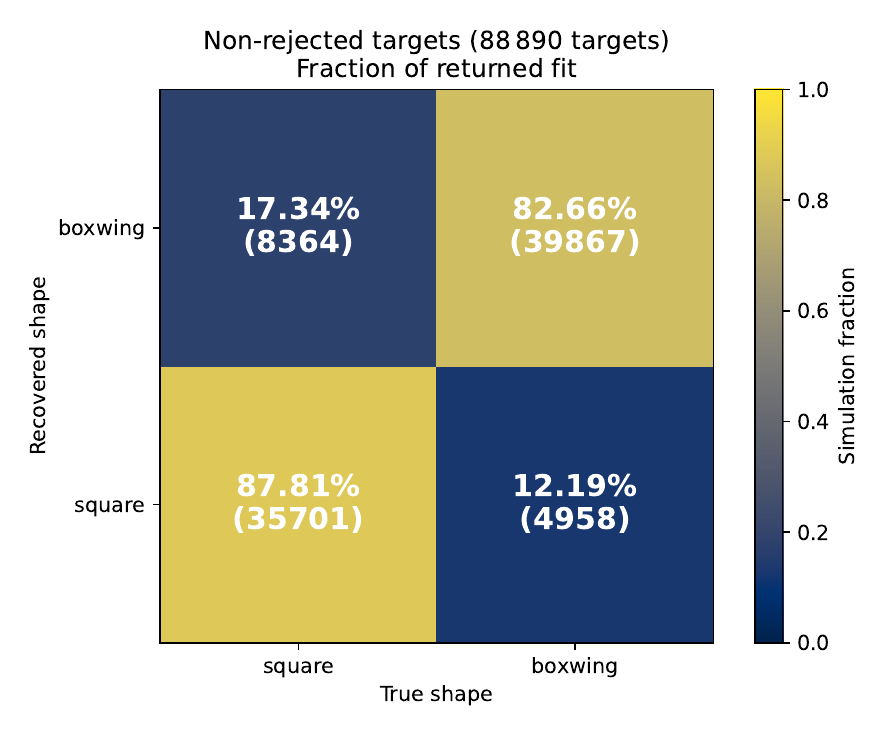}
        \caption{Only including satellites with at least one successful fit}
        \label{fig:accuracy_plot2b}
    \end{subfigure}
    \caption{Satellites separated by their true shape and their shape as returned by our simulation. Percentages are given as a fraction of the number of satellites of each returned classification (i.e. totals are summed horizontally).}
    \label{fig:accuracy_plot2}
\end{figure}

The 100\,000 simulated satellites are given one of three designations depending on their recovered shape. Satellites are designated as correct, when the recovered classification matches the true class; incorrect, when the recovered classification is different to the true class; and rejected, when the data is insufficient to make a robust classification. Figures \ref{fig:accuracy_plot1} and \ref{fig:accuracy_plot2} show the distribution between the satellite designations. In Figure \ref{fig:accuracy_plot1}, percentages are presented as a function of the total number of satellites in each true class, i.e. the percentage tells us what fraction of each satellite class has each recovered designation. In Figure \ref{fig:accuracy_plot2}, percentages are presented as a function of the number of satellites in each recovered class, i.e. the percentage tells us what fraction of each recovered class has each true classification. Figures \ref{fig:accuracy_plot1a} and \ref{fig:accuracy_plot2a} show all 100\,000 satellites, including rejected fits. Figures \ref{fig:accuracy_plot1b} and \ref{fig:accuracy_plot2b} remove rejected satellites from consideration (leaving 88\,890 satellites).

From Figure \ref{fig:accuracy_plot1} we see that square satellites have a slightly worse success rate than boxwing satellites, even when removing rejected satellites. This is likely due to the fact that offset intersection points, caused by the extended exposure time, can look similar to non-offset intersection points located on wings, meaning a misclassification of a square as a boxwing. When intersection points from wings are offset, they can remove the wing effect, or they can look like longer wings, meaning boxwings may be misclassified as squares, or may be correctly classified, simply with longer wings. Additionally, from Figure \ref{fig:accuracy_plot2} we see that square satellites are slightly more likely to be rejected than boxwings, most likely due to the smaller surface area leading to fewer intersection points and thus less robust fits.

\begin{figure*}
    \centering
    \includegraphics[width=2\columnwidth]{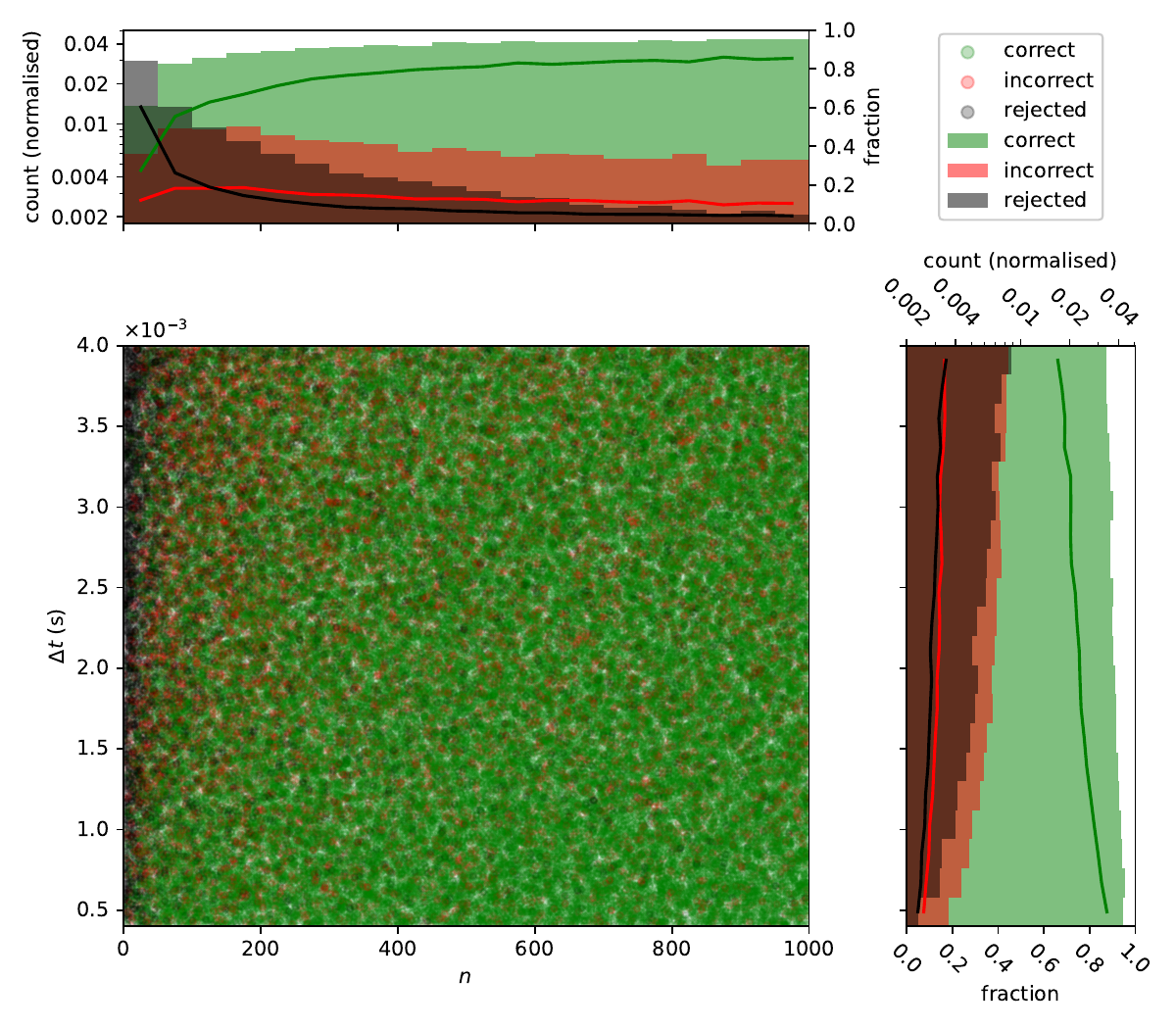}
    \caption{Plot showing the results of our classification simulation as a function of number of stars, $n$, and exposure time, $\Delta t$. The scatter plot shows each satellite, plotted in $n-\Delta t$ parameter space. The points are coloured green, red and black for correct, incorrect and rejected classifications respectively. The marginal histograms show the distributions of each parameter separately. The bars show the (normalised) number of each classification of satellite (correct, incorrect, rejected) in each bin, while the solid lines show the fraction of each bin that are correctly classified, incorrectly classified or rejected, in green, red and black respectively.}
    \label{fig:scatter_hist}
\end{figure*}

Figure \ref{fig:scatter_hist} shows the distribution of successful, unsuccessful and rejected classifications as a function of number of stars, $n$, and exposure time, $\Delta t$. The scatter plot shows the individual simulated satellites, located by their $n$ and $\Delta t$ values and coloured by their classification. The marginal histograms show the results as functions of the two parameters individually, separated into 20 uniform bins. The histogram counts are normalised and, additionally, the histograms include solid line plots, showing the fraction of satellites in each bin with each classification (successful, unsuccessful and rejection).

From this figure we see that the correct classification rate increases with number of stars and decreases with exposure time. Both effects are as expected. More stars along the path of the satellite results in a higher number of stars being occulted by the satellite, and thus a larger number of data points against which to fit a satellite shape. Additionally, a smaller exposure time reduces the error on the location of any particular intersection point, improving the accuracy of the best fits. We also see that the number of rejected satellites (i.e. those without a classification) are inverted with respect to the success rate, decreasing with number of stars and increasing with exposure time.

Looking at the marginal histograms we see that it is the number of stars that has the greater effect on the chance of rejection. This is directly due to the fact that more stars means more data points, and therefore, an increased chance of getting at least one successful fit. For the smallest $n$ bin, rejected satellites make up the largest fraction, with approximately 60\% of satellites being rejected. The fraction of incorrectly classified satellites also increases with exposure time, again due to the decreased accuracy of intersection point placement for longer exposures. The fraction of incorrectly classified satellites actually increases with increasing $n$ to begin with, before decreasing. The increase is simply due to the fact that in the first bin, such a large fraction of targets are rejected, that the fraction of incorrect classifications is forced down. Once the number of rejections falls, we see the expected pattern, which is that the incorrect classification rate falls with increasing $n$, since more stars results in more data points, improving the reliability of shape fits.

Appendix \ref{sec:Resolved occultations only} reproduces Figures \ref{fig:accuracy_plot1}, \ref{fig:accuracy_plot2} and \ref{fig:scatter_hist} under the regime where only fully resolved occultations are considered (as discussed in Sect. \ref{sec:Shape recovery}).

\begin{figure*}
    \centering
    \begin{subfigure}{2\columnwidth}
        \centering
        \includegraphics[width=0.75\columnwidth]{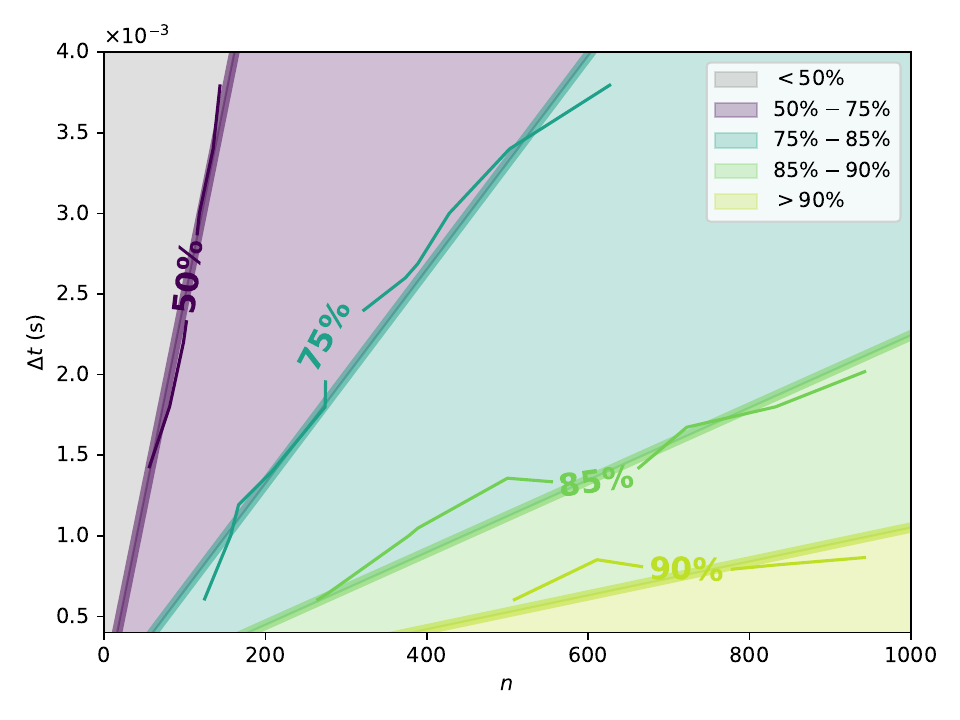}
        \caption{All targets are used. A success is a satellite which is correctly classified and a failure is either an incorrect classification or a rejection. Fit parameters are given in Equation \ref{eq:all_sat_fits}.}
        \label{fig:contoura}
    \end{subfigure}
    \begin{subfigure}{2\columnwidth}
        \centering
        \includegraphics[width=0.75\columnwidth]{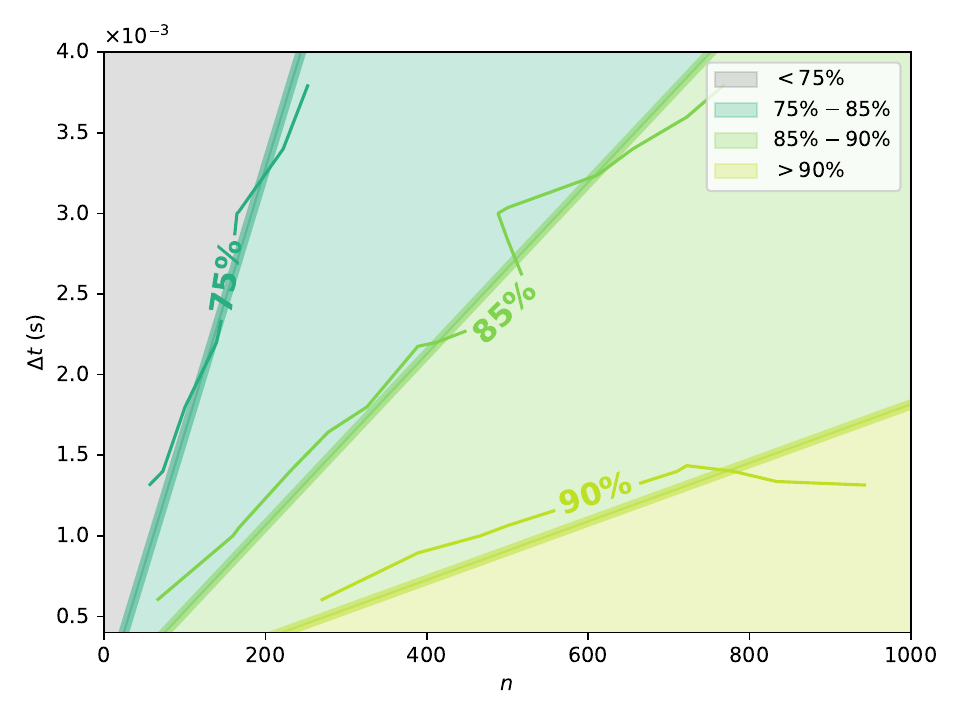}
        \caption{Non-rejected targets are used. A success is a satellite which is correctly classified and a failure is an incorrect classification. Fit parameters are given in Equation \ref{eq:non-rejected_sat_fits}.}
        \label{fig:contourb}
    \end{subfigure}
    \caption{Success rate as function of number of stars, $n$, and exposure time, $\Delta t$. Contours are shown corresponding to a range of success rates, $r$. The thicker lines are fits to the recovered contour coordinates using the parameters given in Equations \ref{eq:all_sat_fits} and \ref{eq:non-rejected_sat_fits}. Shaded regions show the portions of the parameter space between each pair of contour fits.}
    \label{fig:contour}
\end{figure*}

Figure \ref{fig:contour} shows the success rate contours as a function of $n$ and $\Delta t$. Satellites are binned along both axes, and we calculate the success rate within each bin. The thin coloured lines show the success rate contours based on these bins. The solid lines are linear fits to these contours, with the areas between fits being shaded the appropriate colour. We show contours of 50\%, 75\%, 85\% and 90\% (the 50\% contour is not present when rejected satellites are excluded). For a success rate $r$, the exposure time $\Delta t$ as a function of number of stars $n$ $(0 < n \leq 1000)$ is given by:


\begin{equation}
\label{eq:all_sat_fits}
\Delta t=
\begin{dcases}
    2.48\times10^{-5}n, & \text{for } r=50\%\\
    6.63\times10^{-6}n, & \text{for } r=75\%\\
    2.24\times10^{-6}n, & \text{for } r=85\%\\
    1.05\times10^{-6}n, & \text{for } r=90\%\\
\end{dcases}
\end{equation}

when considering all satellites. If rejected satellites are ignored, the fits are given by:


\begin{equation}
\label{eq:non-rejected_sat_fits}
\Delta t=
\begin{dcases}
    1.64\times10^{-5}n, & \text{for } r=75\%\\
    5.32\times10^{-6}n, & \text{for } r=85\%\\
    1.81\times10^{-6}n, & \text{for } r=90\%\\
\end{dcases}
\end{equation}

As expected, and as explained above, success rate increases with number of stars, $n$, and decreases with exposure time, $\Delta t$. In Figure \ref{fig:contoura}, the grey region indicates a success rate of $<50\%$. This is possible, since we include rejected satellites, lowering the success rate below 50\%, the limit that would be expected based on a random classification. This region is not present in Figure \ref{fig:contourb} where rejected satellites are not considered. The success rate is improved when rejected satellites are removed from consideration, simply due to fewer satellites being classified as unsuccessful. This effect is most noticeable for low $n$ and larger $\Delta t$ where the number of rejected satellites is higher (see Figure \ref{fig:scatter_hist}).

Examining the fit parameters given in Equations \ref{eq:all_sat_fits} and \ref{eq:non-rejected_sat_fits} more closely, we find a relation between the fitted parameter values and the relevant success rate, $r$. This relation can be approximated by a quadratic equation of the form $y = ar^2 + br + c$.

Noting that $y$ is the gradient between $n$ and $\Delta t$, we can rearrange this equation to give the following relation:

\begin{equation}
\label{eq:success_rate}
    r = \frac{-b \pm \sqrt{b^2 - 4a(c-\Delta t/n)}}{2a}
\end{equation}

The values of the parameters $a$, $b$ and $c$ depend on whether we are considering all satellites (Equation \ref{eq:all_sat_fits}) or only non-rejected satellites (Equation \ref{eq:non-rejected_sat_fits}). The parameters are given 
Table \ref{tab:contour_fit_params}.



\begin{table}[ht]
\centering
\caption{Contour fit parameters for Equation \ref{eq:success_rate}. Values for all satellite fits and non-rejected satellite fits are given in columns 2 and 3 respectively.}
\label{tab:contour_fit_params}
\begin{tabular}{ccc}
Parameter   &   All satellites    &   Non-rejected satellites   \\ \hline
$a$   &   $9.34\times10^{-9}$   &   $3.09\times10^{-8}$   \\
$b$   &   $-1.90\times10^{-6}$  &   $-6.06\times10^{-6}$  \\
$c$   &   $9.69\times10^{-5}$   &   $2.97\times10^{-4}$
\end{tabular}
\end{table}

Using Equation \ref{eq:success_rate} and Table \ref{tab:contour_fit_params} therefore allows us to predict the success rate $r$ as a function of $n$ and $\Delta t$. The chosen values of $a$, $b$ and $c$ depend on whether or not we are including rejected targets in our sample.

\begin{figure}
    \centering
    \begin{subfigure}{\columnwidth}
        \centering
        \includegraphics[width=\columnwidth]{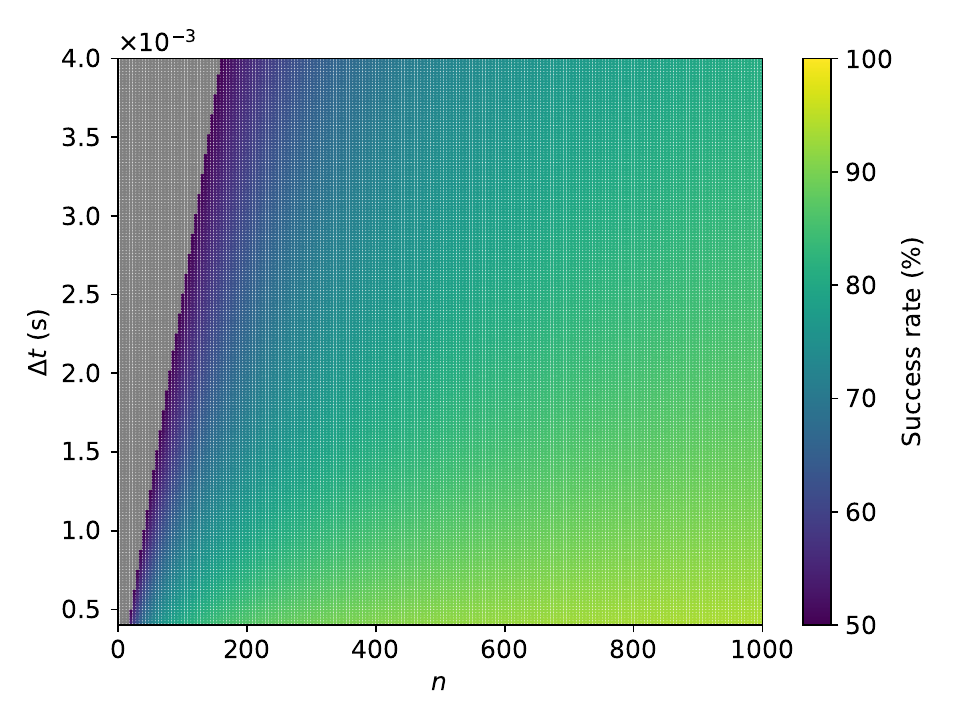}
        \caption{All targets are used. The inferred success rate uses parameters given in 
        Table \ref{tab:contour_fit_params}, column 2.}
        \label{fig:contour_fita}
    \end{subfigure}
    \begin{subfigure}{\columnwidth}
        \centering
        \includegraphics[width=\columnwidth]{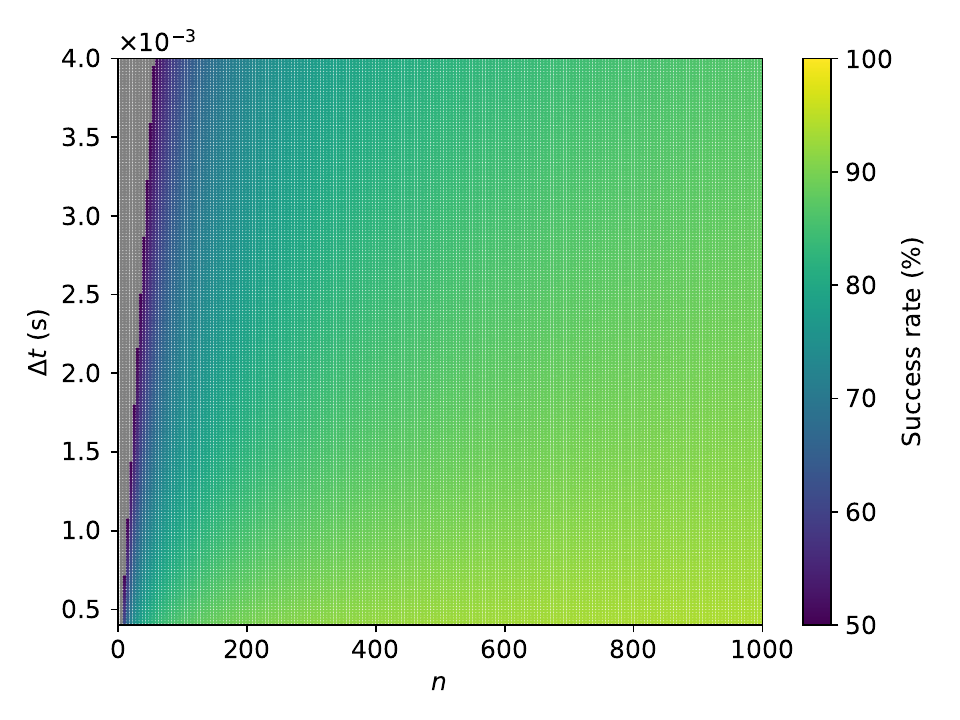}
        \caption{Non-rejected targets are used. The inferred success rate uses parameters given in 
        Table \ref{tab:contour_fit_params}, column 3.}
        \label{fig:contour_fitb}
    \end{subfigure}
    \caption{Inferred success rate as function of number of stars, $n$, and exposure time, $\Delta t$. Success rate is calculated using Equation \ref{eq:success_rate} and parameters from 
    Table \ref{tab:contour_fit_params}. The grey region corresponds to a success rate less than 50\%.}
    \label{fig:contour_fit}
\end{figure}

Figure \ref{fig:contour_fit} shows the result of these predicted success rates. At each point in $n-\Delta t$ space, we calculate $r$ using the above parameters. Points for which $r<50\%$ are marked in grey. Figure \ref{fig:contour_fita} shows the results for all satellites, and Figure \ref{fig:contour_fitb} shows the result when rejected satellites are excluded.

Figure \ref{fig:contour_fit} shows good agreement with Figure \ref{fig:contour}, but allows us to consider each point in $n-\Delta t$ space individually.

\section{Discussion and conclusions}
\label{sec:Discussion and conclusions}


The results presented here suggest that utilising occultations of background stars is a theoretically sound method for successfully classifying LEO satellites. However, due to the difficulty of the technique (specifically the small size of LEO objects and their large speeds) achieving a high classification success rate requires very fast exposure times (hundreds to thousands of frames per second) and dense stellar fields (hundreds of stars along the LEO object's path), putting the practical limits of the technique beyond the capabilities of current systems. For a given success rate, there is shown to be an approximately linear relationship between the two parameters, meaning that a lack in one can be compensated for by the other (i.e., a larger exposure time would require a greater stellar density, equivalently, a reduced stellar density would require correspondingly shorter exposure times). However, even with this trade off, the requirements are substantial.
The high densities required, especially near the top end of our range of $n$ values, are only readily available in limited areas of the sky, and possibly not even then when requiring such short exposure times. Increasing the length of the observed arc would help reduce the required density (as mentioned in Sect. \ref{sec:Simulation parameters}), but care would need to be taken to avoid a significant increase to the chance of the satellite cross section changing during the observation window (discussed further below). Increasing the length of the observed arc could be done in a number of ways. If restricted to sidereally tracked observations the amount of sky visible to the telescope would have to be increased, either by using a telescope with a larger FoV or by using multiple FoVs positioned along the track of the satellite (this could require one or more separate telescopes depending on the speed of the satellite motion and telescope movement capabilities). Another method would be to track the satellite directly. This would allow observations to be made for a longer arc by a single telescope, but would involve more moving parts and thus potentially introduce more sources of error and noise. Characterisation success is highest in regions of high stellar density, i.e. when observations are taken as a target passes across the Milky Way. Therefore, an effective strategy is to concentrate observations on these regions of the sky.


The results presented above rely on a number of assumptions about the observations and targets. The first of these is the assumption that photometry of the occulted stars can be recovered successfully and accurately. Accurate photometry is required to correctly identify occultation events and locate the data points used to classify the targets. To be able to realistically identify occultation events the technique requires a telescope capable of measuring photometry of the affected stars, to the level associated with an occultation, using the chosen exposure time. At LEO, assuming a velocity of 7.5\,km/s (as discussed in Sect. \ref{sec:Simulation parameters}), a target of size 1\,m will occult a point source for $1.\dot{3}\times10^{-4}$\,s. Taking an exposure time of $2.0\times10^{-3}$\,s means that such an occultation will cause a flux drop of $6.\dot{6}\%$ (an equal size target at GEO, with a velocity of 3.0\,km/s, will occult a point source for $3.\dot{3}\times10^{-4}$\,s and cause a flux drop of $16.\dot{6}\%$ in a $2.0\times10^{-3}$\,s exposure; \citealt{2017amos.confE..80S}). Thus, to use the technique presented here, we would require a telescope capable of conducting $\sim$ percent level photometry at a rate of hundreds to thousands of frames per second (longer exposures would require a corresponding increase in photometric precision).

Simulations and calculations above assume observations of targets at or near zenith. As the target moves closer to the horizon, the satellite-observer distance increases, up to a factor of $\sim2.1$ at an elevation of $20\degree$ above the horizon (taken as the limiting case, see \citealt{COOKE_2023}). The angular size of a satellite scales by the inverse of this distance. Additionally, as the target moves to lower elevations, its apparent angular velocity is reduced, as described in Sect. \ref{sec:Simulation parameters}. These two factors do not cancel exactly, thus we find that for lower elevation observations, occultations are longer, increasing by a up to a factor of $\sim1.7$ at $20\degree$ above the horizon. The result of this is that occultations are easier to detect closer to the horizon, however the sky area traversed in the same time is reduced, increasing the required stellar density.


The second assumption made is that the shape of the satellite, as seen by the observer, remains constant over the length of the observation. For sufficiently short observations, i.e., a relatively small section of a full pass, this may be true for a spin-stabilised satellite (the condition would hold true for longer arcs if the satellite were in a higher altitude orbit, for example at GEO). For longer observations, or for tumbling or rotating targets, this assumption breaks down. The classification of a target assumed to be rotating would require more data, and to consider that data with a temporal component, since the cross section presented to an observer would change as a function of time. Considering rotation is beyond the scope of this manuscript but is a topic of potential future study. Using this technique to classify active satellites however, should be less affected by this assumption, since active satellites are more likely to be spin-stabilised (though, for LEO, we must still assume short arcs). During a satellites orbit the orientation of its solar panels (the wings of the boxwing shape) will change in orientation to track the sun. This will cause the wings to appear to increase and decrease in width from the observers point of view. To allay this effect, the recovered success rates above are averaged over a range of wing widths (given in Sect. \ref{sec:Simulation parameters}), equivalent to using a range of solar panel orientation angles. However, characterisation will be easier when the wings appearing wider, thus timing observations based on expected solar panel orientation may be a judicious approach.


It should also be noted that we have assumed a single value for LEO satellite height, and thus velocity. In reality, height and velocity are part of a non-uniform distribution \citep{COOKE_2023}. Lower-altitude orbits have larger velocities, but will also have larger apparent sizes, with the reverse being true for higher-altitude orbits. The result is, that for LEO height orbits, occultation duration is broadly unchanged as a function of height. At significantly higher orbits, such as those at GEO, the overall result is a longer occultation duration \citep{2022A&A...667A..45G}. Additionally, GEO satellites are generally physically larger than those at LEO resulting in more easily detected occultations. Since the goal of this manuscript is simply to find some limits on a proof-of-concept technique we simplify by assuming a single-valued distribution, under the assumption that it has minimal effect at LEO, but in practise this is not the case.

When measuring the photometry, we assume that the position of the background stars is known to a very high degree of precision. These positions are used to infer the location of star/satellite intersection points, relative to the centre of the satellite. In practice, the errors on these measurements will be small, but non-zero, due to database uncertainties (in the case of particularly faint stars) and seeing effects. The requirement to use many stars in order to reach sufficient densities will increase the impact of these effects. We assume these sources of error to be marginal compared to some of the assumptions we have already made and thus do not implement them here. Additionally, these effects are likely to be telescope, site and night specific, and the goal of this paper is a generalised exploration of the method and its limitations. A preferred alternative approach, to avoid seeing related location errors, is to use the photometry to determine which stars are occulted when, and then extract the locations from a database of sufficient accuracy (for example, Gaia DR3, which has precise positions and proper motions of more than a billion stars \citep{2023A&A...674A...1G}). This would mean that the telescope needs only to be able to identify which star, from a known database, is being occulted, instead of having to accurately locate each star, possibly reducing pixel scale and other observational requirements.

It may be possible to observe an occulting target with multiple telescopes simultaneously. For neighbouring telescopes, observing concurrently may lead to an improvement in photometric precision by combining observations, but would then multiply the required telescope time. Additionally, combining observations could help to limit possible false positives (an issue we don't touch upon here). For more distantly separated observers, the parallax effect provides slightly different sight-lines on the target. The range of angles could potentially add extra information to the occultations data and aid in the shape recovery efforts. Similar parallax effects have been used in the detection and characterisation of asteroids \citep{2021PASP..133k4401G,2024A&A...683A.122G}. We mention this idea as a potential future extension to the project, but don't discuss it further here.

Finally, we conduct this simulation on satellites for which the orbit is known, but the size and shape are unknown. As this is a proof-of-concept study we operate under the assumption the the orbit of the satellite has been identified using more standard techniques that struggle with identifying size/shape information (as discussed in Sect. \ref{sec:Introduction}). As a follow-up study we hope to be able to relax this assumption, using the occultations themselves to detect unknown satellites, and extrapolate an orbit, while also allowing for classification.

The method described in this paper is built upon the use of a sCMOS system to gather optical observations of a stellar field, and then extract occultation data from the resulting photometry. An alternative approach, worthy of mention, would be to use a neuromorphic, or event-based, camera in place of a sCMOS. Neuromorphic cameras are sensors which react to changes in brightness \citep{5f53418988f64a2eae4d0e7cfe97fd55}. Each pixel independently measures a brightness value and are only triggered when they detect a sufficiently large change. A neuromorphic camera, under noiseless conditions, observing a stellar field and tracking sidereally, would only return data when the field changes, i.e., when a star dims due to an occultation (in reality the camera would have to be capable of rejecting dimming events caused by seeing effects and other false positives). A system such as this could be used, alongside a well characterised stellar field, to immediately identify the time and location of occultations, without having to worry about exposure time effects. The main benefit of such a method would be a precise timestamp corresponding to the start of occultation events, but to be fully successful the neuromorphic camera would have to have sufficient sensitivity and resolution, with limited pixel latency, to correctly identify occultations and locate them accurately in space and time. The use of neuromorphic cameras in the field of space domain awareness is still relatively novel, and in most cases has yet to reach the performance of optimised sCMOS systems \citep{2023amos.conf..137M,2023amos.conf..142M}, thus we don't explore the idea beyond mentioning it here, but the technique offers a potentially complimentary methodology for future research in this area.

Short of direct, resolved observations utilising AO or ISAR (each with their own complexities and limitations; described in Sect. \ref{sec:Introduction}), occultations of background stars offer us one of the only methods to directly determine the size and shape of a LEO satellite. As discussed in Sect. \ref{sec:Introduction}, other methods may measure proxies for size (i.e. RCS or magnitude) but not size directly. Additionally, these alternative methods are sensitive to satellite reflectivity and composition, and are compromised by deliberate attempts to reduce the detectability of a satellite. Characterisation through occultations of background stars is independent of either composition or reflectivity, and is unaffected by satellite-level obfuscation efforts.

A potential future direction for this research would be to attempt to reduce some of the assumptions stated here. For example, lifting the assumption of zero rotation, or even using observations of a full arc, would drastically change the satellite cross section available to the observer. This would in turn make classification much more complex, but this aspect could be simplified, attempting a simpler characterisation such as maximum satellite dimension. In this case we would remove all attempted recovery of rotation or shape and measure simply the separation of inferred intersection points from occultation data, while accounting for changing viewing angle or rotation. This approach would be simpler in terms of characterisation, but more physically realistic than the one presented here, requiring fewer assumptions. An alternative approach would be to restrict our simulation to higher altitude targets, for example, focusing on GEO satellites (similar suggestions are made by \cite{2022A&A...667A..45G}). This would also enable us to reduce our assumptions but would restrict the accessible parameter space to targets further from Earth. However, this change of parameter space may potentially make it more feasible to combine a simulation with true observations, enabling an empirical verification of the methodology.

In conclusion, we have presented a proof-of-concept for a LEO satellite classification technique using occultations of background stars. The results shown here prove that the technique is theoretically sound and place some limits on the requirements to use the method practically. These limits are sufficiently extreme as to be generally beyond the scope of current optical observation systems, but we verify the method as a potential tool for future observatories.

\section*{Acknowledgements}
\label{sec:Acknowledgements}

BFC acknowledges support from a Science and Technology Facilities Council CLASP award (grant ST/V002279/1) and from the Defence Science and Technology Laboratory (UK).
JAB acknowledges support from the Science and Technology Facilities Council (grant ST/Y50998X/1).
The authors thank the anonymous reviewers for their comments which have helped to improve this manuscript.

For the purpose of open access, the author has applied a Creative Commons Attribution (CC-BY) licence to any Author Accepted Manuscript version arising from this submission.

\section*{Data Availability}
\label{sec:Data Availability}


Relevant data are freely available under a CC-BY-4.0 license at \href{https://github.com/BenCooke95/Occultation\_simulation}{https://github.com/BenCooke95/Occultation\_simulation}. Further details may be granted upon reasonable request to the corresponding author.

\bibliographystyle{rasti}
\bibliography{manuscript_revised_clean} 

\begin{thebibliography}{37}
\expandafter\ifx\csname natexlab\endcsname\relax\def\natexlab#1{#1}\fi

\bibitem[Airey et~al.(2025)Airey, Chote, Blake, Cooke, McCormac, Allen, MacManus, Pollacco, Shrive, \& West]{AIREY20255757}
Airey, R.~J., Chote, P., Blake, J.~A., Cooke, B.~F., McCormac, J., Allen, P., MacManus, A., Pollacco, D., Shrive, B., \& West, R., 2025.
\newblock A comprehensive survey of the geo-belt using simultaneous four-colour observations with sting, {\it Advances in Space Research\/}, {\bf 75}(7), 5757--5780.

\bibitem[{Anger} et~al.(2018){Anger}, {Jirousek}, {Dill}, {Schreiber}, \& {Peichl}]{2018SPIE10633E..0LA}
{Anger}, S., {Jirousek}, M., {Dill}, S., {Schreiber}, E., \& {Peichl}, M., 2018.
\newblock {Imaging of satellites in space (IoSiS): challenges in image processing of ground-based high-resolution ISAR data}, in {\em Radar Sensor Technology XXII\/}, vol. 10633 of {\bf Society of Photo-Optical Instrumentation Engineers (SPIE) Conference Series}, p. 106330L.

\bibitem[{Blake}(2022)]{2022A&G....63.2.14B}
{Blake}, J.~A., 2022.
\newblock {Looking out for a sustainable space}, {\it Astronomy and Geophysics\/}, {\bf 63}(2), 2.14--2.20.

\bibitem[{Cooke} et~al.(2023){Cooke}, {Chote}, {Pollacco}, {West}, {Blake}, {McCormac}, {Airey}, \& {Shrive}]{COOKE_2023}
{Cooke}, B.~F., {Chote}, P., {Pollacco}, D., {West}, R., {Blake}, J.~A., {McCormac}, J., {Airey}, R., \& {Shrive}, B., 2023.
\newblock {Simulated recovery of LEO objects using sCMOS blind stacking}, {\it Advances in Space Research\/}, {\bf 72}(4), 907--921.

\bibitem[Cooke et~al.(2024)Cooke, Blake, Chote, McCormac, \& Pollacco]{COOKE_2024}
Cooke, B.~F., Blake, J.~A., Chote, P., McCormac, J., \& Pollacco, D., 2024.
\newblock {Predicting RSO populations using a neighbouring orbits technique}, {\it RAS Techniques and Instruments\/}, p. rzae034.

\bibitem[{Curzi} et~al.(2020){Curzi}, {Modenini}, \& {Tortora}]{2020Aeros...7..133C}
{Curzi}, G., {Modenini}, D., \& {Tortora}, P., 2020.
\newblock {Large Constellations of Small Satellites: A Survey of Near Future Challenges and Missions}, {\it Aerospace\/}, {\bf 7}(9), 133.

\bibitem[Emery \& Camps(2017)]{EMERY2017291}
Emery, W. \& Camps, A., 2017.
\newblock Chapter 5 - {Radar}, in {\em Introduction to Satellite Remote Sensing\/}, pp. 291--453, eds Emery, W. \& Camps, A., Elsevier.

\bibitem[{Gaia Collaboration} et~al.(2023){Gaia Collaboration}, {Vallenari}, {Brown}, {Prusti}, {de Bruijne}, {Arenou}, {Babusiaux}, {Biermann}, {Creevey}, {Ducourant}, {Evans}, {Eyer}, {Guerra}, {Hutton}, {Jordi}, {Klioner}, {Lammers}, {Lindegren}, {Luri}, {Mignard}, {Panem}, {Pourbaix}, {Randich}, {Sartoretti}, {Soubiran}, {Tanga}, {Walton}, {Bailer-Jones}, {Bastian}, {Drimmel}, {Jansen}, {Katz}, {Lattanzi}, {van Leeuwen}, {Bakker}, {Cacciari}, {Casta{\~n}eda}, {De Angeli}, {Fabricius}, {Fouesneau}, {Fr{\'e}mat}, {Galluccio}, {Guerrier}, {Heiter}, {Masana}, {Messineo}, {Mowlavi}, {Nicolas}, {Nienartowicz}, {Pailler}, {Panuzzo}, {Riclet}, {Roux}, {Seabroke}, {Sordo}, {Th{\'e}venin}, {Gracia-Abril}, {Portell}, {Teyssier}, {Altmann}, {Andrae}, {Audard}, {Bellas-Velidis}, {Benson}, {Berthier}, {Blomme}, {Burgess}, {Busonero}, {Busso}, {C{\'a}novas}, {Carry}, {Cellino}, {Cheek}, {Clementini}, {Damerdji}, {Davidson}, {de Teodoro}, {Nu{\~n}ez Campos}, {Delchambre}, {Dell'Oro}, {Esquej},
  {Fern{\'a}ndez-Hern{\'a}ndez}, {Fraile}, {Garabato}, {Garc{\'\i}a-Lario}, {Gosset}, {Haigron}, {Halbwachs}, {Hambly}, {Harrison}, {Hern{\'a}ndez}, {Hestroffer}, {Hodgkin}, {Holl}, {Jan{\ss}en}, {Jevardat de Fombelle}, {Jordan}, {Krone-Martins}, {Lanzafame}, {L{\"o}ffler}, {Marchal}, {Marrese}, {Moitinho}, {Muinonen}, {Osborne}, {Pancino}, {Pauwels}, {Recio-Blanco}, {Reyl{\'e}}, {Riello}, {Rimoldini}, {Roegiers}, {Rybizki}, {Sarro}, {Siopis}, {Smith}, {Sozzetti}, {Utrilla}, {van Leeuwen}, {Abbas}, {{\'A}brah{\'a}m}, {Abreu Aramburu}, {Aerts}, {Aguado}, {Ajaj}, {Aldea-Montero}, {Altavilla}, {{\'A}lvarez}, {Alves}, {Anders}, {Anderson}, {Anglada Varela}, {Antoja}, {Baines}, {Baker}, {Balaguer-N{\'u}{\~n}ez}, {Balbinot}, {Balog}, {Barache}, {Barbato}, {Barros}, {Barstow}, {Bartolom{\'e}}, {Bassilana}, {Bauchet}, {Becciani}, {Bellazzini}, {Berihuete}, {Bernet}, {Bertone}, {Bianchi}, {Binnenfeld}, {Blanco-Cuaresma}, {Blazere}, {Boch}, {Bombrun}, {Bossini}, {Bouquillon}, {Bragaglia}, {Bramante}, {Breedt},
  {Bressan}, {Brouillet}, {Brugaletta}, {Bucciarelli}, {Burlacu}, {Butkevich}, {Buzzi}, {Caffau}, {Cancelliere}, {Cantat-Gaudin}, {Carballo}, {Carlucci}, {Carnerero}, {Carrasco}, {Casamiquela}, {Castellani}, {Castro-Ginard}, {Chaoul}, {Charlot}, {Chemin}, {Chiaramida}, {Chiavassa}, {Chornay}, {Comoretto}, {Contursi}, {Cooper}, {Cornez}, {Cowell}, {Crifo}, {Cropper}, {Crosta}, {Crowley}, {Dafonte}, {Dapergolas}, {David}, {David}, {de Laverny}, {De Luise}, \& {De March}]{2023A&A...674A...1G}
{Gaia Collaboration}, {Vallenari}, A., {Brown}, A.~G.~A., {Prusti}, T., {de Bruijne}, J.~H.~J., {Arenou}, F., {Babusiaux}, C., {Biermann}, M., {Creevey}, O.~L., {Ducourant}, C., {Evans}, D.~W., {Eyer}, L., {Guerra}, R., {Hutton}, A., {Jordi}, C., {Klioner}, S.~A., {Lammers}, U.~L., {Lindegren}, L., {Luri}, X., {Mignard}, F., {Panem}, C., {Pourbaix}, D., {Randich}, S., {Sartoretti}, P., {Soubiran}, C., {Tanga}, P., {Walton}, N.~A., {Bailer-Jones}, C.~A.~L., {Bastian}, U., {Drimmel}, R., {Jansen}, F., {Katz}, D., {Lattanzi}, M.~G., {van Leeuwen}, F., {Bakker}, J., {Cacciari}, C., {Casta{\~n}eda}, J., {De Angeli}, F., {Fabricius}, C., {Fouesneau}, M., {Fr{\'e}mat}, Y., {Galluccio}, L., {Guerrier}, A., {Heiter}, U., {Masana}, E., {Messineo}, R., {Mowlavi}, N., {Nicolas}, C., {Nienartowicz}, K., {Pailler}, F., {Panuzzo}, P., {Riclet}, F., {Roux}, W., {Seabroke}, G.~M., {Sordo}, R., {Th{\'e}venin}, F., {Gracia-Abril}, G., {Portell}, J., {Teyssier}, D., {Altmann}, M., {Andrae}, R., {Audard}, M., {Bellas-Velidis},
  I., {Benson}, K., {Berthier}, J., {Blomme}, R., {Burgess}, P.~W., {Busonero}, D., {Busso}, G., {C{\'a}novas}, H., {Carry}, B., {Cellino}, A., {Cheek}, N., {Clementini}, G., {Damerdji}, Y., {Davidson}, M., {de Teodoro}, P., {Nu{\~n}ez Campos}, M., {Delchambre}, L., {Dell'Oro}, A., {Esquej}, P., {Fern{\'a}ndez-Hern{\'a}ndez}, J., {Fraile}, E., {Garabato}, D., {Garc{\'\i}a-Lario}, P., {Gosset}, E., {Haigron}, R., {Halbwachs}, J.~L., {Hambly}, N.~C., {Harrison}, D.~L., {Hern{\'a}ndez}, J., {Hestroffer}, D., {Hodgkin}, S.~T., {Holl}, B., {Jan{\ss}en}, K., {Jevardat de Fombelle}, G., {Jordan}, S., {Krone-Martins}, A., {Lanzafame}, A.~C., {L{\"o}ffler}, W., {Marchal}, O., {Marrese}, P.~M., {Moitinho}, A., {Muinonen}, K., {Osborne}, P., {Pancino}, E., {Pauwels}, T., {Recio-Blanco}, A., {Reyl{\'e}}, C., {Riello}, M., {Rimoldini}, L., {Roegiers}, T., {Rybizki}, J., {Sarro}, L.~M., {Siopis}, C., {Smith}, M., {Sozzetti}, A., {Utrilla}, E., {van Leeuwen}, M., {Abbas}, U., {{\'A}brah{\'a}m}, P., {Abreu Aramburu}, A.,
  {Aerts}, C., {Aguado}, J.~J., {Ajaj}, M., {Aldea-Montero}, F., {Altavilla}, G., {{\'A}lvarez}, M.~A., {Alves}, J., {Anders}, F., {Anderson}, R.~I., {Anglada Varela}, E., {Antoja}, T., {Baines}, D., {Baker}, S.~G., {Balaguer-N{\'u}{\~n}ez}, L., {Balbinot}, E., {Balog}, Z., {Barache}, C., {Barbato}, D., {Barros}, M., {Barstow}, M.~A., {Bartolom{\'e}}, S., {Bassilana}, J.~L., {Bauchet}, N., {Becciani}, U., {Bellazzini}, M., {Berihuete}, A., {Bernet}, M., {Bertone}, S., {Bianchi}, L., {Binnenfeld}, A., {Blanco-Cuaresma}, S., {Blazere}, A., {Boch}, T., {Bombrun}, A., {Bossini}, D., {Bouquillon}, S., {Bragaglia}, A., {Bramante}, L., {Breedt}, E., {Bressan}, A., {Brouillet}, N., {Brugaletta}, E., {Bucciarelli}, B., {Burlacu}, A., {Butkevich}, A.~G., {Buzzi}, R., {Caffau}, E., {Cancelliere}, R., {Cantat-Gaudin}, T., {Carballo}, R., {Carlucci}, T., {Carnerero}, M.~I., {Carrasco}, J.~M., {Casamiquela}, L., {Castellani}, M., {Castro-Ginard}, A., {Chaoul}, L., {Charlot}, P., {Chemin}, L., {Chiaramida}, V., {Chiavassa},
  A., {Chornay}, N., {Comoretto}, G., {Contursi}, G., {Cooper}, W.~J., {Cornez}, T., {Cowell}, S., {Crifo}, F., {Cropper}, M., {Crosta}, M., {Crowley}, C., {Dafonte}, C., {Dapergolas}, A., {David}, M., {David}, P., {de Laverny}, P., {De Luise}, F., \& {De March}, R., 2023.
\newblock {Gaia Data Release 3. Summary of the content and survey properties}, {\it \aap\/}, {\bf 674}, A1.

\bibitem[{Garc{\'\i}a-Mart{\'\i}n} et~al.(2024){Garc{\'\i}a-Mart{\'\i}n}, {Kruk}, {Popescu}, {Mer{\'\i}n}, {Stapelfeldt}, {Evans}, {Carry}, \& {Thomson}]{2024A&A...683A.122G}
{Garc{\'\i}a-Mart{\'\i}n}, P., {Kruk}, S., {Popescu}, M., {Mer{\'\i}n}, B., {Stapelfeldt}, K.~R., {Evans}, R.~W., {Carry}, B., \& {Thomson}, R., 2024.
\newblock {Hubble Asteroid Hunter. III. Physical properties of newly found asteroids}, {\it \aap\/}, {\bf 683}, A122.

\bibitem[{Giovinazzi} et~al.(2021){Giovinazzi}, {Blake}, \& {Bernardinelli}]{2021PASP..133k4401G}
{Giovinazzi}, M.~R., {Blake}, C.~H., \& {Bernardinelli}, P.~H., 2021.
\newblock {Enhancing Ground-based Observations of Trans-Neptunian Objects Using a Single-epoch Parallax Measurement from L2}, {\it \pasp\/}, {\bf 133}(1029), 114401.

\bibitem[{Gomes-J{\'u}nior} et~al.(2021){Gomes-J{\'u}nior}, {Morgado}, {Rossi}, {Boufleur}, {Rommel}, \& {Huarca}]{2021ascl.soft08025G}
{Gomes-J{\'u}nior}, A.~R., {Morgado}, B.~E., {Rossi}, G.~B., {Boufleur}, R.~C., {Rommel}, F.~L., \& {Huarca}, M.~B., 2021.
\newblock {SORA: Stellar Occultation Reduction Analysis}, {\it Astrophysics Source Code Library\/}, ascl:2108.025.

\bibitem[Gronchi et~al.(2015)Gronchi, Dimare, Bracali~Cioci, \& Ma]{10.1093/mnras/stv1010}
Gronchi, G.~F., Dimare, L., Bracali~Cioci, D., \& Ma, H., 2015.
\newblock {On the computation of preliminary orbits for Earth satellites with radar observations}, {\it Monthly Notices of the Royal Astronomical Society\/}, {\bf 451}(2), 1883--1891.

\bibitem[{Groot}(2022)]{2022A&A...667A..45G}
{Groot}, P.~J., 2022.
\newblock {Satellite shadows through stellar occultations}, {\it \aap\/}, {\bf 667}, A45.

\bibitem[{IADC}(2007)]{LEO}
{IADC}, 2007.
\newblock {IADC Space Debris Mitigation Guidelines, IADC-02-01}, Tech. rep., INTER-AGENCY SPACE DEBRIS COORDINATION COMMITTEE, Steering Group and Working Group 4.

\bibitem[{Jennings-Bramly} \& {Maxey}(2023)]{2023amos.conf...34J}
{Jennings-Bramly}, T. \& {Maxey}, J., 2023.
\newblock {An End-to-End Signal Processing Chain for Low Earth Orbit Inverse Synthetic Aperture Radar Space Object Imaging}, in {\em Proceedings of the Advanced Maui Optical and Space Surveillance (AMOS) Technologies Conference\/}, p.~34.

\bibitem[Kirkland et~al.(2023)Kirkland, Clemente, Macdonald, {Di Caterina}, \& Meoni]{5f53418988f64a2eae4d0e7cfe97fd55}
Kirkland, P., Clemente, C., Macdonald, M., {Di Caterina}, G., \& Meoni, G., 2023.
\newblock Neuromorphic sensing and processing for space domain awareness, in {\em 2023 International Geoscience and Remote Sensing Symposium, IGARSS2023 - Proceedings\/}, IEEE.

\bibitem[{Marcireau} et~al.(2023){Marcireau}, {Afshar}, {Ralph}, {Jones}, \& {Cohen}]{2023amos.conf..137M}
{Marcireau}, A., {Afshar}, S., {Ralph}, N.~O., {Jones}, I., \& {Cohen}, G., 2023.
\newblock {Binocular Telescope for Neuromorphic Space Situational Awareness}, in {\em Proceedings of the Advanced Maui Optical and Space Surveillance (AMOS) Technologies Conference\/}, p. 137.

\bibitem[Martorella(2014)]{MARTORELLA2014987}
Martorella, M., 2014.
\newblock Chapter 19 - introduction to inverse synthetic aperture radar, in {\em Academic Press Library in Signal Processing: Volume 2\/}, vol.~2 of {\bf Academic Press Library in Signal Processing}, pp. 987--1042, eds Sidiropoulos, N.~D., Gini, F., Chellappa, R., \& Theodoridis, S., Elsevier.

\bibitem[{McReynolds} et~al.(2023){McReynolds}, {Graca}, {Oliver}, {Nishiguchi}, \& {Delbruck}]{2023amos.conf..142M}
{McReynolds}, B., {Graca}, R., {Oliver}, R., {Nishiguchi}, M., \& {Delbruck}, T., 2023.
\newblock {Demystifying Event-based Sensor Biasing to Optimize Signal to Noise for Space Domain Awareness}, in {\em Proceedings of the Advanced Maui Optical and Space Surveillance (AMOS) Technologies Conference\/}, p. 142.

\bibitem[Muciaccia et~al.(2024)Muciaccia, Facchini, Montaruli, Purpura, Detomaso, Colombo, Massari, {Di Lizia}, {Di Cecco}, Salotti, \& Bianchi]{MUCIACCIA2024143}
Muciaccia, A., Facchini, L., Montaruli, M.~F., Purpura, G., Detomaso, R., Colombo, C., Massari, M., {Di Lizia}, P., {Di Cecco}, A., Salotti, L., \& Bianchi, G., 2024.
\newblock Radar observation and reconstruction of cosmos 1408 fragmentation, {\it Journal of Space Safety Engineering\/}, {\bf 11}(1), 143--149.

\bibitem[{Muelhaupt} et~al.(2019){Muelhaupt}, {Sorge}, {Morin}, \& {Wilson}]{2019JSSE....6...80M}
{Muelhaupt}, T.~J., {Sorge}, M.~E., {Morin}, J., \& {Wilson}, R.~S., 2019.
\newblock {Space traffic management in the new space era}, {\it The Journal of Space Safety Engineering\/}, {\bf 6}(2), 80--87.

\bibitem[{Muruganandan} et~al.(2023){Muruganandan}, {Lambert}, {Liu}, {Clare}, \& {Weddell}]{2023amos.conf..148M}
{Muruganandan}, V.~A., {Lambert}, A., {Liu}, J., {Clare}, R., \& {Weddell}, S., 2023.
\newblock {Partial Image Reconstruction of an Artificial Satellite in Real Time using Background Natural Stars.}, in {\em Proceedings of the Advanced Maui Optical and Space Surveillance (AMOS) Technologies Conference\/}, p. 148.

\bibitem[{Olivieri} \& {Francesconi}(2020)]{2020AdSpR..65..351O}
{Olivieri}, L. \& {Francesconi}, A., 2020.
\newblock {Large constellations assessment and optimization in LEO space debris environment}, {\it Advances in Space Research\/}, {\bf 65}(1), 351--363.

\bibitem[{Pardini} \& {Anselmo}(2023)]{2023AcAau.210..465P}
{Pardini}, C. \& {Anselmo}, L., 2023.
\newblock {The short-term effects of the Cosmos 1408 fragmentation on neighboring inhabited space stations and large constellations}, {\it Acta Astronautica\/}, {\bf 210}, 465--473.

\bibitem[Petit et~al.(2020)Petit, Mugnier, Bonnefois, Conan, Fusco, Levraud, Meimon, Michau, Montri, Vedrenne, Velluet, \& F{\'e}tick]{petit:hal-03051923}
Petit, C., Mugnier, L., Bonnefois, A., Conan, J.-M., Fusco, T., Levraud, N., Meimon, S., Michau, V., Montri, J., Vedrenne, N., Velluet, M.-T., \& F{\'e}tick, R., 2020.
\newblock {LEO satellite imaging with adaptive optics and marginalized blind deconvolution}, in {\em {21st AMOS Advanced Maui Optical and Space Surveillance Technologies Conference}\/}, Virtuel, United States.

\bibitem[{Schulz}(1993)]{1993JOSAA..10.1064S}
{Schulz}, T.~J., 1993.
\newblock {Multiframe blind deconvolution of astronomical images.}, {\it Journal of the Optical Society of America A\/}, {\bf 10}(5), 1064--1073.

\bibitem[{Selvan} et~al.(2023){Selvan}, {Siemuri}, {Prol}, {V{\"a}lisuo}, {Bhuiyan}, \& {Kuusniemi}]{2023GPSS...27..178S}
{Selvan}, K., {Siemuri}, A., {Prol}, F.~S., {V{\"a}lisuo}, P., {Bhuiyan}, M. Z.~H., \& {Kuusniemi}, H., 2023.
\newblock {Precise orbit determination of LEO satellites: a systematic review}, {\it GPS Solutions\/}, {\bf 27}(4), 178.

\bibitem[{Sheppard} et~al.(2017){Sheppard}, {Douglas}, {Hunt}, \& {Todoki}]{2017amos.confE..80S}
{Sheppard}, D.~G., {Douglas}, D.~M., {Hunt}, B.~R., \& {Todoki}, A., 2017.
\newblock {Recent Developments in Shadow Imaging}, in {\em Advanced Maui Optical and Space Surveillance (AMOS) Technologies Conference\/}, p.~80.

\bibitem[{Suthakar} et~al.(2023){Suthakar}, {Sanvido}, {Qashoa}, \& {Lee}]{2023Senso..23.9668S}
{Suthakar}, V., {Sanvido}, A.~A., {Qashoa}, R., \& {Lee}, R. S.~K., 2023.
\newblock {Comparative Analysis of Resident Space Object (RSO) Detection Methods}, {\it Sensors\/}, {\bf 23}(24), 9668.

\bibitem[Tokunaga(2014)]{TOKUNAGA20141089}
Tokunaga, A.~T., 2014.
\newblock Chapter 51 - new generation ground-based optical/infrared telescopes, in {\em Encyclopedia of the Solar System (Third Edition)\/}, pp. 1089--1105, eds Spohn, T., Breuer, D., \& Johnson, T.~V., Elsevier, Boston, third edition edn.

\bibitem[{Tregloan-Reed} et~al.(2020){Tregloan-Reed}, {Otarola}, {Ortiz}, {Molina}, {Anais}, {Gonz{\'a}lez}, {Colque}, \& {Unda-Sanzana}]{2020A&A...637L...1T}
{Tregloan-Reed}, J., {Otarola}, A., {Ortiz}, E., {Molina}, V., {Anais}, J., {Gonz{\'a}lez}, R., {Colque}, J.~P., \& {Unda-Sanzana}, E., 2020.
\newblock {First observations and magnitude measurement of Starlink's Darksat}, {\it \aap\/}, {\bf 637}, L1.

\bibitem[{Tregloan-Reed} et~al.(2021){Tregloan-Reed}, {Otarola}, {Unda-Sanzana}, {Haeussler}, {Gaete}, {Colque}, {Gonz{\'a}lez-Fern{\'a}ndez}, {Anais}, {Molina}, {Gonz{\'a}lez}, {Ortiz}, {Mieske}, {Brillant}, \& {Anderson}]{2021A&A...647A..54T}
{Tregloan-Reed}, J., {Otarola}, A., {Unda-Sanzana}, E., {Haeussler}, B., {Gaete}, F., {Colque}, J.~P., {Gonz{\'a}lez-Fern{\'a}ndez}, C., {Anais}, J., {Molina}, V., {Gonz{\'a}lez}, R., {Ortiz}, E., {Mieske}, S., {Brillant}, S., \& {Anderson}, J.~P., 2021.
\newblock {Optical-to-NIR magnitude measurements of the Starlink LEO Darksat satellite and effectiveness of the darkening treatment}, {\it \aap\/}, {\bf 647}, A54.

\bibitem[{Werth} et~al.(2019){Werth}, {Calef}, {Roe}, \& {Conti}]{2019amos.confE...5W}
{Werth}, M., {Calef}, B., {Roe}, K., \& {Conti}, A., 2019.
\newblock {Multi-Frame Blind Deconvolution Accelerated with Graphical Processing Units (GPUs)}, in {\em Advanced Maui Optical and Space Surveillance Technologies Conference\/}, p.~5.

\bibitem[{Xu} et~al.(2019){Xu}, {Kennedy}, \& {Stansbery}]{2019LPICo2109.6164X}
{Xu}, Y.~L., {Kennedy}, T.~F., \& {Stansbery}, E.~G., 2019.
\newblock {Radar Cross Section of Orbital Debris Objects}, in {\em First International Orbital Debris Conference\/}, vol. 2109 of {\bf LPI Contributions}, p. 6164.

\bibitem[Zhao et~al.(2016)Zhao, Zhang, Yu, \& Mao]{ZHAO20162269}
Zhao, X.-F., Zhang, H.-Y., Yu, Y., \& Mao, Y.-D., 2016.
\newblock Multicolor photometry of geosynchronous satellites and application on feature recognition, {\it Advances in Space Research\/}, {\bf 58}(11), 2269--2279.

\bibitem[Zhi et~al.(2024)Zhi, Jiang, \& Wang]{10.1093/mnras/stae693}
Zhi, H., Jiang, X., \& Wang, J., 2024.
\newblock {Multicolour photometry of LEO mega-constellations Starlink and OneWeb}, {\it Monthly Notices of the Royal Astronomical Society\/}, {\bf 530}(4), 5006--5015.

\bibitem[{Zhu} et~al.(2023){Zhu}, {Zheng}, {Luo}, {Chen}, {Wang}, {Mao}, {Wu}, \& {Wang}]{2023AdSpR..72.2064Z}
{Zhu}, Z.-H., {Zheng}, J.-H., {Luo}, H., {Chen}, G.-P., {Wang}, W., {Mao}, Y.-D., {Wu}, W.-T., \& {Wang}, K.-P., 2023.
\newblock {CMOS-based observation of Resident Space Objects using short-exposure stacking mode}, {\it Advances in Space Research\/}, {\bf 72}(6), 2064--2077.

\end{thebibliography}



\appendix

\section{Resolved occultations only}
\label{sec:Resolved occultations only}

In this section we consider the effect on the successful classification rate of satellites if only fully resolved occultations are included. As discussed in Sects. \ref{sec:Photometry generation} and \ref{sec:Shape recovery} occultation times can only be fully resolved if the occultation itself occurs across one or more exposure boundaries. If this is the case, the occultation time is fully resolved, if not, only the occultation duration is recovered. The full resolution, or otherwise, of individual occultations can be inferred from lightcurves thus can be known for real data. Here, we reject intersection points which have an inexact temporal location.

\begin{figure}
    \centering
    \begin{subfigure}{\columnwidth}
        \centering
        \includegraphics[width=\columnwidth]{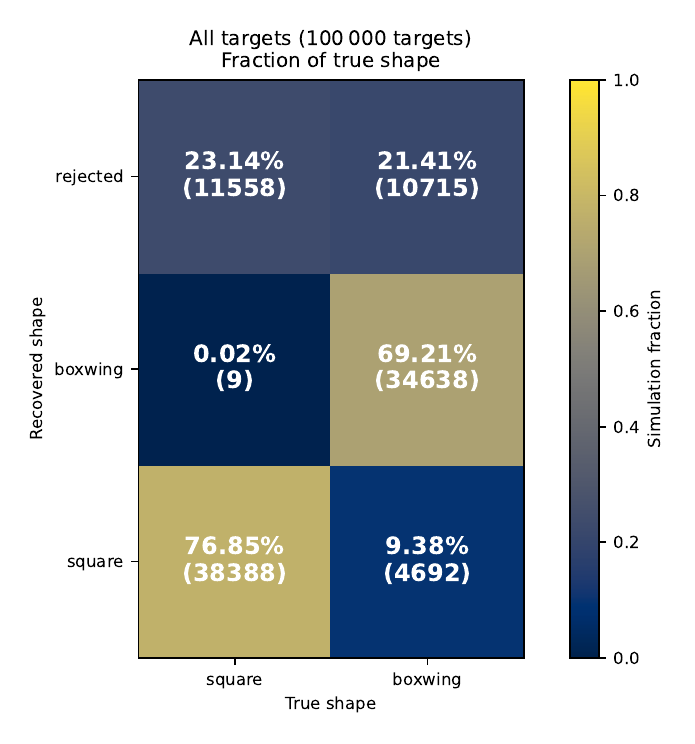}
        \caption{Including satellites with no successful fits}
        \label{fig:accuracy_plot1a_new}
    \end{subfigure}
    \begin{subfigure}{\columnwidth}
        \centering
        \includegraphics[width=\columnwidth]{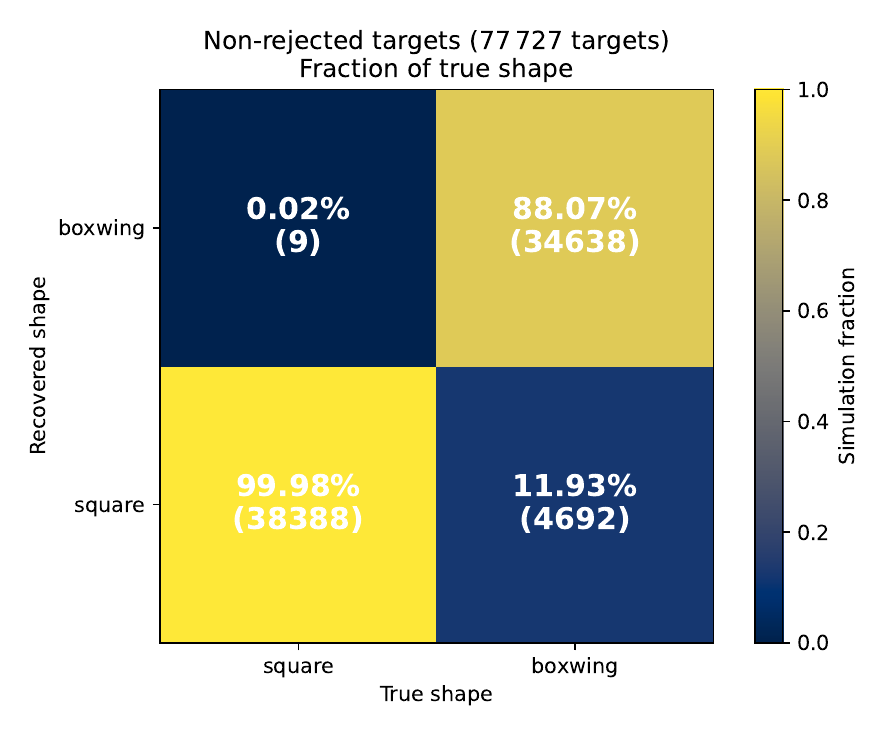}
        \caption{Only including satellites with at least one successful fit}
        \label{fig:accuracy_plot1b_new}
    \end{subfigure}
    \caption{Satellites separated by their true shape and their shape as returned by our simulation. Percentages are given as a fraction of the number of satellites of each true classification (i.e. totals are summed vertically). Results are obtained using only fully resolved occultations.}
    \label{fig:accuracy_plot1_new}
\end{figure}

\begin{figure}
    \centering
    \begin{subfigure}{\columnwidth}
        \centering
        \includegraphics[width=\columnwidth]{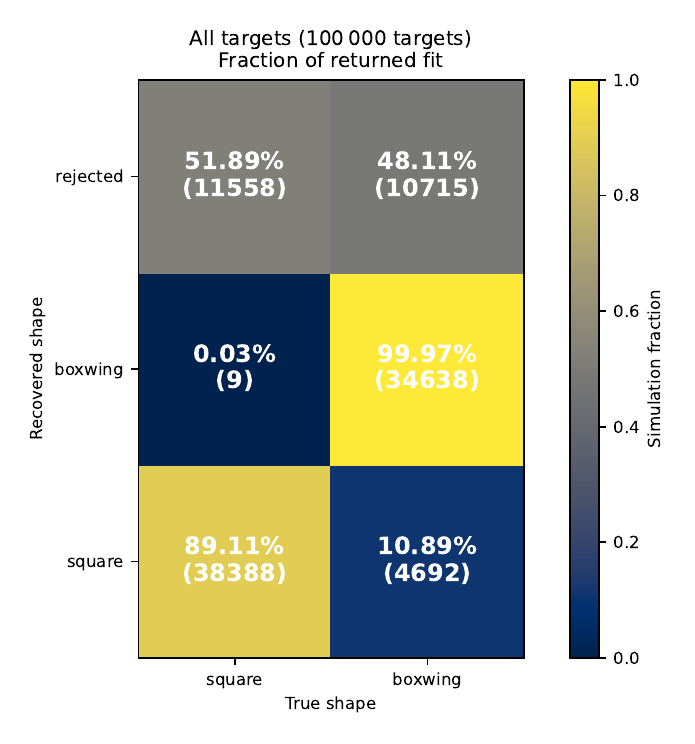}
        \caption{Including satellites with no successful fits}
        \label{fig:accuracy_plot2a_new}
    \end{subfigure}
    \begin{subfigure}{\columnwidth}
        \centering
        \includegraphics[width=\columnwidth]{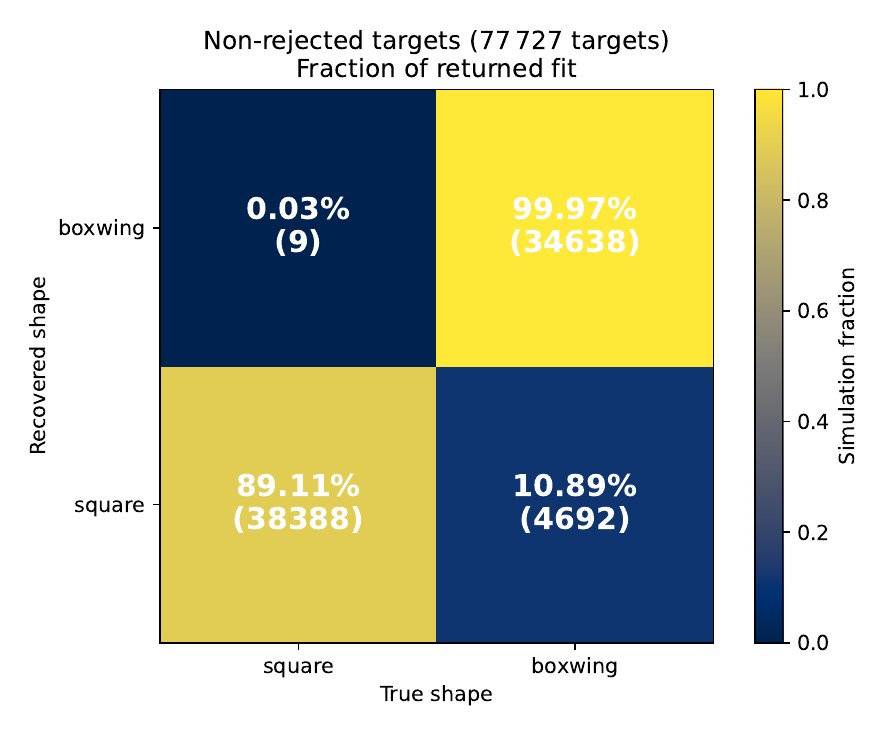}
        \caption{Only including satellites with at least one successful fit}
        \label{fig:accuracy_plot2b_new}
    \end{subfigure}
    \caption{Satellites separated by their true shape and their shape as returned by our simulation. Percentages are given as a fraction of the number of satellites of each returned classification (i.e. totals are summed horizontally). Results are obtained using only fully resolved occultations.}
    \label{fig:accuracy_plot2_new}
\end{figure}

The result of this is that there is reduced scatter of the recovered intersection points around the true edges of the satellite (as seen in Figure \ref{fig:best_fits}), this converse however, is that approximate data points are lost. Data points are more likely to be lost when caused by shorter occultations as a result of thinner structures, such as wings, thus boxwing and square satellites are not affected equally. Figures \ref{fig:accuracy_plot1_new} and \ref{fig:accuracy_plot2_new} reproduce Figures \ref{fig:accuracy_plot1} and \ref{fig:accuracy_plot2} when only fully resolved data points are considered.

From these figures we notice a number of differences when compared with the equivalent figures in the main body of this paper. The first significant change is that the fraction of rejected targets has increased from 11\% to 22\%. This is a direct result of fewer data points being included in the shape recovery steps. It is easier to recover a best fitting shape from noisy data than from no data at all. When looking at results as a fraction of returned fit we see that almost 100\% of boxwing fits are correct. For a target to return a boxwing fit it must show visible wing structures in the data, highly unlikely to occur for a square satellite when only fully resolved intersection points are included. This is higher than the equivalent value when using all points as noisy points can appear as wing structures. The fraction of square fits which are found to be correct is 89\%, comparable to the equivalent value found when using all data points. This value is lower than the boxwing value since it is easier to miss wing structures than it is to find them where they are not present. This value is mainly unchanged from the all points regime since square satellites are less affected by noise as they have fewer thin features.

When including all satellites, classified and rejected, we find that 69\% of boxwing satellites are correctly classified, lower than the 79\% found when all points are considered. This is a combination of two effects. Firstly, is that a higher fraction of boxwing satellites are rejected due to insufficient data. Secondly, the non-rejected boxwings have an equivalent success rate as in the all points regime, this is due to the fact that in the original regime, aligned points are already weighted more highly, thus removing noisy points has limited effect. The combination of these effects is a reduced overall success rate for boxwing satellites.

For square satellites, we find the overall success rate to be 77\%, higher than the 72\% found when considering all points. Again, this is caused by a combination of two effects. Firstly, the recurring effect of more rejected satellites due to fewer data points. In this case however, the second effect is that non-rejected squares have an improved success rate compared to the all points regime ($\sim$100\% vs 81\%), this is due to the fact that, without noisy points, it is unlikely to fit wing structures where they do not exist, whereas noisy points can be misinterpreted. The combination of these effects is an enhanced overall success rate for square satellites.

Overall, we find that the success rate in the regime where only fully resolved data are considered is 73\%, lower than the 76\% success rate found when all data points are included. This is a consequence of the two major effects; that there are fewer data points per satellite, but that the accepted data points are less noisy.

\begin{figure*}
    \centering
    \includegraphics[width=2\columnwidth]{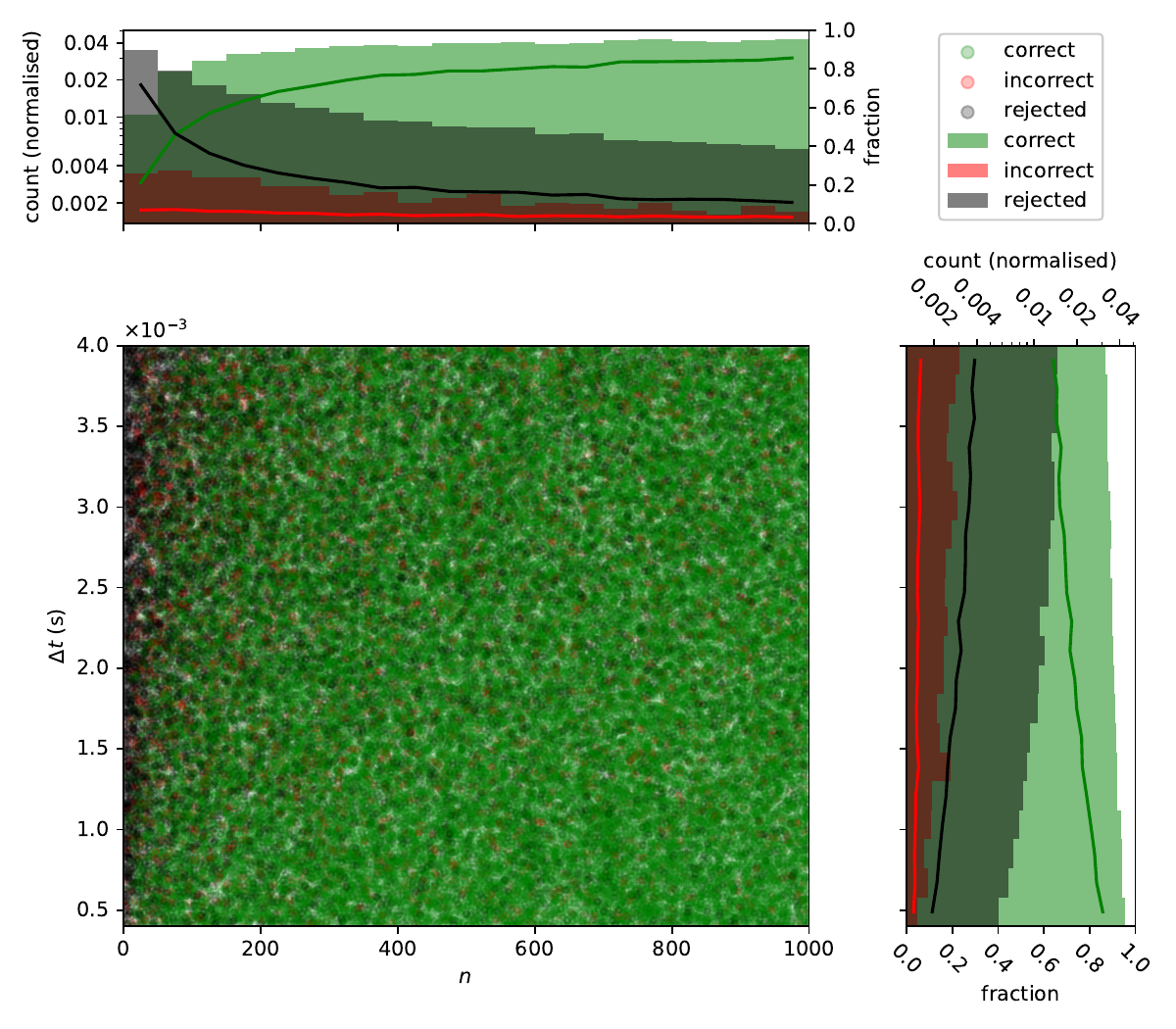}
    \caption{Plot showing the results of our classification simulation as a function of number of stars, $n$, and exposure time, $\Delta t$. The scatter plot shows each satellite, plotted in $n-\Delta t$ parameter space. The points are coloured green, red and black for correct, incorrect and rejected classifications respectively. The marginal histograms show the distributions of each parameter separately. The bars show the (normalised) number of each classification of satellite (correct, incorrect, rejected) in each bin, while the solid lines show the fraction of each bin that are correctly classified, incorrectly classified or rejected, in green, red and black respectively. Results are obtained using only fully resolved occultations.}
    \label{fig:scatter_hist_new}
\end{figure*}

Figure \ref{fig:scatter_hist_new} reproduces Figure \ref{fig:scatter_hist} in the regime where only fully resolved data points are considered.

From this figure we see that the general patterns and trends match those seen in the corresponding all points data regime, the key of these being that the success rate increases with $n$ and decreases with $\Delta t$. The main contrast between the two regimes is the increased number of rejected targets when considering fully resolved occultations only. At all points in the parameter space, the fraction of rejected points is higher than in the all points regime, and additionally, is always higher than the fraction of incorrectly classified points. This means that, unlike in the all points regime, a target is always more likely to be rejected than incorrectly classified. This follows from the fact that significant amounts of data are rejected, but where sufficient data is present to make a classification, the data has minimal noise, allowing for an enhanced success rate (without accounting for differences between the two classes of shape as discussed above). The number of stars, $n$, is seen to have a stronger effect on recovery than $\Delta t$, due to the fact that even long exposures will result in some resolved data points (assuming sufficient $n$) where an occultation crosses an exposure boundary, whereas a low number of stars will result in a limited amount of data even when $\Delta t$ is low. As discussed in the above paragraphs, the overall success rate in this regime is lower than that found in the all data points regime.

The approach of considering fully resolved data only has advantages and disadvantages compared to the method discussed in the body of this paper, that is, considering all data points together. The significant differences are an increased fraction of rejected satellite fits, due to a limited amount of data being included in the shape recovery and analysis steps, and a reduction in noise of the surviving data. Since we find this more limited regime to produce a slightly weaker overall successful classification rate (73\% vs 76\%), we leave further consideration of this regime in this work, however we do present it here as an alternative method that may be preferred depending on the exact goals of any future classification studies.






\bsp	
\label{lastpage}
\end{document}